\documentclass[12pt]{article}
\newif\iffigs\figstrue
\usepackage{latexsym}
\iffigs
  \input{epsf}
\else
\fi
\font\tenmsbm=msbm10 scaled 1200
\font\sevenmsbm=msbm9
\newfam\msbmfam
\textfont\msbmfam=\tenmsbm
\scriptfont\msbmfam=\sevenmsbm

\makeatletter
\@addtoreset{equation}{section}
\makeatother

\newcommand{\eqn}[1]{(\ref{#1})}
\newcommand{\ft}[2]{{\textstyle\frac{#1}{#2}}}
\newsavebox{\uuunit}
\sbox{\uuunit}
	{\setlength{\unitlength}{0.825em}
	 \begin{picture}(0.6,0.7)
		\thinlines
		\put(0,0){\line(1,0){0.5}}
		\put(0.15,0){\line(0,1){0.7}}
		\put(0.35,0){\line(0,1){0.8}}
	   \multiput(0.3,0.8)(-0.04,-0.02){10}{\rule{0.5pt}{0.5pt}}
	 \end {picture}}

\newsavebox{\bobox}
\sbox{\bobox}
	{\setlength{\unitlength}{0.825em}
	 \begin{picture}(0.6,0.7)
		\thinlines
		\put(0,0){\line(1,0){0.7}}
		\put(0,0){\line(0,1){0.7}}
		\put(0.7,0){\line(0,1){0.7}}
		\put(0,0.7){\line(1,0){0.7}}
	   \multiput(0,0)(0.05,0.05){14}{\rule{0.4pt}{0.4pt}}
	   \multiput(0,0.65)(0.05,-0.05){13}{\rule{0.4pt}{0.4pt}}
	 \end {picture}}
\newcommand {\xbox}{\mathord{\!\usebox{\bobox}}\,}

\def\IP{\relax{\rm I\kern-.18em P}}

\font\cmss=cmss10 \font\cmsss=cmss10 at 7pt

\def\inbar{\vrule height1.5ex width.4pt depth0pt}
\def\IC{\relax\,\hbox{$\inbar\kern-.3em{\rm C}$}}
\def\IG{\relax\,\hbox{$\inbar\kern-.3em{\rm G}$}}
\def\IB{\relax{\rm I\kern-.18em B}}
\def\ID{\relax{\rm I\kern-.18em D}}
\def\IL{\relax{\rm I\kern-.18em L}}
\def\IF{\relax{\rm I\kern-.18em F}}
\def\IH{\relax{\rm I\kern-.18em H}}
\def\II{\relax{\rm I\kern-.17em I}}
\def\IN{\relax{\rm I\kern-.18em N}}
\def\IM{\relax{\rm I\kern-.18em M}}
\def\IP{\relax{\rm I\kern-.18em P}}
\def\IQ{\relax\,\hbox{$\inbar\kern-.3em{\rm Q}$}}
\def\bfzero{\relax\,\hbox{$\inbar\kern-.3em{\rm 0}$}}
\def\IK{\relax{\rm I\kern-.18em K}}
\def\IG{\relax\,\hbox{$\inbar\kern-.3em{\rm G}$}}
 \font\cmss=cmss10 \font\cmsss=cmss10 at 7pt
\def\IR{\relax{\rm I\kern-.18em R}}
\def\ZZ{\relax\ifmmode\mathchoice
{\hbox{\cmss Z\kern-.4em Z}}{\hbox{\cmss Z\kern-.4em Z}}
{\lower.9pt\hbox{\cmsss Z\kern-.4em Z}}
{\lower1.2pt\hbox{\cmsss Z\kern-.4em Z}}\else{\cmss Z\kern-.4em
Z}\fi}
\def\bfone{\relax{\rm 1\kern-.35em 1}}

 \def\cB{{\cal B}}
\def\cC{{\cal C}}

\def\cL{{\cal L}}

\def\cR{{\cal R}} 
 
\def\IE{\relax{{\rm I\kern-.18em E}}}

\def\IGam{\relax{{\rm I}\kern-.18em \Gamma}}

\def\bet{\begin{tabular}}
\def\eet{\end{tabular}}

\def\a{\alpha}
\def\b{\beta}
\def\l{\lambda}

\def\g{\gamma}
\def\d{\delta}

\def\r{\rho}
\def\s{\sigma}

\def\e{\epsilon}

\begin{document}
\begin{titlepage}
\begin{flushright}
hep-th/0012159 \\
DFTT 50/2000 \\
\end{flushright}
\vskip 2cm
\begin{center}
{{\Large \bf M -theory on $AdS_4\times Q^{111}$:\\} \vskip 0.2cm
{\Large \bf  the  complete $Osp(2\vert 4)\times SU(2)\times SU(2)\times SU(2)$ spectrum \\
\vskip 0.2cm from harmonic analysis \hskip 0.2cm $^\dagger$
}}\\
\vfill
\end{center}

\centerline{\bf Paolo Merlatti} \centerline{\sl Dipartimento di Fisica Teorica,
Universit\`a di Torino} \centerline{\sl and I.N.F.N., sezione di Torino} \centerline{\sl
via P. Giuria 1, I-10125 Torino, Italy} \centerline{\sl Paolo.Merlatti@to.infn.it}

\vfill
\begin{abstract}
{In this paper by means of harmonic analysis we derive the
complete spectrum of $Osp\left( 2|4\right)\times SU(2)\times
SU(2)\times SU(2)$ multiplets that one obtains compactifying
$D=11$ supergravity on the homogeneous
space $Q^{111}$.\\
In particular we analyze the structure of the short multiplets and compare them with the
corresponding composite operators of the ${\cal N}=2$ conformal field theory dual to such
a compactification, found in a previous publication. We get complete agreement between the
quantum numbers of the supergravity multiplets on one side and those of the conformal
operators on the other side, confirming the structure of the
conjectured $SCFT$. \\
 However the determination of the actual
spectrum by harmonic analysis teaches us a lot more: indeed we find out which multiplets
are present for each representation of the isometry group, how many they are, the exact
values of the hypercharge and of the ``energy'' for each multiplet.}
\end{abstract}
\vspace{2mm} \vfill \hrule width 3.cm {\footnotesize
 $^ \dagger $ \hskip 0.1cm Research partially supported by the EC  RTN programme
HPRN-CT-2000-00131}
\end{titlepage}

\section{Introduction}

The basic principle of the $AdS/CFT$ correspondence \cite{adscft,gub,hol} states that
every consistent $M$-theory or type II background with metric $AdS_{p+2}\times M^{d-p-2}$
in $d$-dimensions, where $M^{d-p-2}$ is an Einstein manifold, is associated with a
conformal quantum field theory living on the boundary of $AdS_{p+2}$. The background is
typically generated by the near horizon geometry of a set of p-branes and the boundary
conformal field theory is identified with the IR limit of the gauge theory
living on the world-volume of the p-branes. \\
In this paper we focus on the case $p=2$ when $M$ is a coset manifold $G/H$ with ${\cal
N}=2$ supersymmetry. In this case one remarkable example with a non-trivial smooth
manifold $M^{7}=M^{111}$ was found in \cite{SCQ,m111}. Some general properties and the
complete spectrum of the $M^{111}$ compactification have been discussed in \cite{m111},
while in \cite{SCQ} the associated superconformal theory that matches supergravity
predictions has been identified.\\ For the manifold $M^{7}=Q^{111}$ an ansatz for the
superconformal field theory was given in \cite{SCQ}\footnote{A first attempt at the
derivation of such a superconformal field theory dual of $Q^{111}$ was put forward in
\cite{tatar}}. In this paper we study $D=11$ supergravity compactified on $Q^{111}$. So
in this case we reverse the usual procedure and we rather use the SCFT side to make
prediction about the Kaluza-Klein spectrum. We show that all the predictions are
completely verified.\\
Actually this paper constitutes a step of a wider project which deals with the specific
case of $D=11$ supergravity compactified on internal homogeneous seven-manifolds:
$M^7=(G/H)^7$. Many examples were studied in the old days of Kaluza Klein theories
\cite{rub} (see \cite{14,15,16,17,18,19,20,21,22} for the cases $S^7$ and squashed $S^7$,
see \cite{23,24,25,26,27,28} for the case $M^{p,q,r}$, see \cite{29,30} for the case
$Q^{111}$, see \cite{22,25,uni,32,33} for general methods of harmonic analysis in
compactification and the structure of $Osp(N|4)$ supermultiplets, for an updated recent
review see \cite{LeonG}, and finally see \cite{34} for a
complete classification of $G/H$ compactifications.)\\
This paper is organized as follows. In section two we briefly review the essential
features of $Q^{111}$ geometry. In section three, using standard techniques of harmonic
analysis, we compute the spectrum of the seven-dimensional Laplace-Beltrami operators
that we need to calculate the complete mass-spectrum of the four-dimensional fields.
Finally in section four we organize all the fields in multiplets of the relevant
superalgebra.

\section{The differential geometry of $Q^{111}$}\label{geometry}
Let us briefly review the essential features of $Q^{111}$
geometry. For notations and basic lore we mainly refer to
\cite{24,25,libro}.
\par
The homogeneous space $Q^{111}$ is the quotient of $G=SU(2) \times SU(2)' \times SU(2)''
\times U(1)$ by the action of its subgroup $H=U(1)' \times U(1)'' \times U(1)^{'''}$,
where the embedding of $H$ in $G$ is defined as follows. The $H$ subgroup is generated by
$Z^{(1)}\ Z^{(2)}$ and $Z^{(3)}$, being three independent linear combinations of the four
abelian generators of $G$.
 Using  the Pauli matrices for
$SU\left(2\right)$ and being $K$ the generator of the $U(1) \in G$ these four generators
are $J_3=-\frac{i}{2}\sigma_{3},\ J_{3'}=-\frac{i}{2}\sigma_{3}',\
J_{3''}-\frac{i}{2}\sigma_{3}''$ and $K$. The three linear combinations $Z^{(1)}$,
$Z^{(2)}$ and $Z^{(3)}$ are defined as a basis for the orthogonal complement of the
generator
\begin{equation} \label{Z}
Z=-\frac{1}{\sqrt{3}}\left( \frac{i}{2}\sigma_{3} +
\frac{i}{2}\sigma_{3}' + \frac{i}{2}\sigma_{3}'' \right)
\end{equation}
 A basis for the three abelian generators of $H$ is then given by
\begin{eqnarray}
Z^{(1)} &=& \frac{1}{\sqrt{3}}\left( -\frac{i}{2}\sigma_{3} + \frac{i}{2}\sigma_{3}'
\right)
 \label{s1}\\ Z^{(2)}&=& \frac{1}{\sqrt{3}}\left( -\frac{i}{2}\sigma_{3} - \frac{i}{2}\sigma_{3}'
 +2\frac{i}{2}\sigma_{3}'' \right) \label{s2}\\
Z^{(3)} &=& -\frac{1}{\sqrt{3}}\left( \frac{i}{2}\sigma_{3} + \frac{i}{2}\sigma_{3}' +
\frac{i}{2}\sigma_{3}'' + 3K\right)\label{s3}
\end{eqnarray}

From an explicit parametrization of the coset $G/H$ we can construct the left-invariant one-forms on $G/H$ as:
\begin{equation}
\Omega(y)=L^{-1}(y){\rm d}L(y)=\Omega^{\Lambda}(y)T_{\Lambda}\,,
\end{equation}
which satisfy the Maurer-Cartan equations
\begin{equation}\label{Omega}
{\rm d}\Omega^{\Lambda}+\ft{1}{2}\cC^{\Lambda}_{\ \Sigma\Pi}
\Omega^{\Sigma}\wedge\Omega^{\Pi}=0
\end{equation}
with the structure constants of $G$:
\begin{equation}
\left[T_{\Sigma},T_{\Pi}\right]=\cC^{\Lambda}_{\ \Sigma\Pi}T_{\Lambda}\,.
\end{equation}
The one-forms $\Omega^{\Lambda}$ can be separated into a set
$\{\Omega^H\}$ corresponding to the generators of the subalgebra
$\IH$ and a set $\{\Omega^{\a}\}$ corresponding to the coset
generators. These latter can be identified with the $SU(2)\times
SU(2)'\times SU(2)''$ invariant seven-vielbeins on $G/H$:
\begin{eqnarray}
\cB^{\a} \equiv (\cB^i, \cB^{i'}, \cB^{i''}, \cB^{z}),\nonumber\\
\left\{\begin{array}{ccl}\label{B}
\cB^i & = & \ft{\sqrt{2}}{8}\Omega^i,\\
\cB^{i'} & = & \ft{\sqrt{2}}{8}\Omega^{i'},\\
\cB^{i''} & = & \ft{\sqrt{2}}{8}\Omega^{i''}\\
\cB^z & = & \ft{1}{8}(\Omega^{0}+\Omega^{0'}+\Omega^{0''})=\ft{\sqrt{3}}{8}\Omega^{z},

\end{array}\right.
\end{eqnarray}
where the multiplicative coefficients in front of the vielbeins
have been properly chosen to let the metric on $G/H$ be Einstein.

The invariant forms $\Omega^H$ are:
\begin{equation}
\left\{\begin{array}{ccl}

\Omega^{Z^{(1)}} & = & \ft{\sqrt{3}}{2}(\Omega^0-\Omega^{0'}),\\
\Omega^{Z^{(2)}} & = & \ft{\sqrt{3}}{6}(\Omega^{0}+\Omega^{0'}-2\Omega^{0''}).
\end{array}\right.
\end{equation}
\par
The spin-connection $\cB^{\a}_{\ \b}$ is easily determined from
the vielbeins $\cB^{\a}$ by imposing vanishing torsion:
\begin{equation}
{\rm d}\cB^{\a}-\cB^{\a}_{\ \b}\wedge\cB^{\b}=0,
\end{equation}
\begin{equation}
\left\{
\begin{array}{ccl}
\cB^{ij} &=& \ft{1}{\sqrt{3}}\e^{ij}\left(\ft{2\sqrt{3}}{3}\cB^z+\cB^{z^{(1)}}+\cB^{z^{(2)}}\right),\\
\cB^{i'j'} &=& \frac{1}{\sqrt{3}}\e^{i'j'}\left(\ft{2\sqrt{3}}{3}\cB^z-\cB^{z^{(1)}}+\cB^{z^{(2)}}\right),\\
\cB^{i''j''} &=&  \frac{1}{\sqrt{3}}\e^{i''j''}\left(\ft{2\sqrt{3}}{3}\cB^z-2\cB^{z^{(2)}}\right),\\
\cB^{iz} &=& -2\e^{i}_{j}\cB^{j},\\
\cB^{i'z} &=& -2\e^{i'}_{j'}\cB^{j'},\\
\cB^{i''z} &=& -2\e^{i''}_{j''}\cB^{j''}.
\end{array}\right.
\end{equation}
\section{Harmonic analysis on $Q^{111}$}
In this section we summarize the essential ideas concerning the
techniques of harmonic analysis on homogeneous seven--manifolds
originally developed in \cite{25,uni}. These techniques are the
basic ingredient of our calculations and in the present summary
we present them already applied to the specific case  of the
$Q^{111}$ manifold.
\par
The essential goal of harmonic analysis is that of translating a differential equation
problem into a linear algebraic one, by means of group theory. In the present case, the
differential equations to solve are the linearized field equations of Kaluza Klein
supergravity, whose typical form is:
\begin{equation}
\left( \Box_x^{[J_1 J_2]}+\xbox_y^{[\l_1\l_2\l_3]}\right)
\Phi_{[\l_1\l_2\l_3]}^{[J_1 J_2]}(x,y)=0,
\end{equation}
where $\Phi_{[\l_1\l_2\l_3]}^{[J_1 J_2]}(x,y)$ is a field transforming
in the irreducible representations $[J_1 J_2]$ of $SO(3,2)$
and  $[\l_1\l_2\l_3]$ of $SO(7)$, and depends both on the
coordinates $x$ of Anti-de Sitter space and on the coordinates $y$
of $G/H$.
$\Box_x^{[J_1 J_2]}$ is the kinetic operator for a field of spin
$[J_1 J_2]$ in four dimension while $\xbox_y^{[\l_1\l_2\l_3]}$ is
the kinetic operator for a field of spin $[\l_1\l_2\l_3]$ in
seven dimensions.
\par
Now, the harmonics constitute a complete set of functions for the
expansion of any $SO(7)$-irreducible field over $G/H$, ${\cal
Y}^i_{[\l_1\l_2\l_3]}(y)$. But their most important property is
that they transform irreducibly under $G$, the group of
isometries of the coset space. This group acts on ${\cal
Y}^i_{[\l_1\l_2\l_3]}(y)$ through the so-called covariant Lie
derivative (see eq.(2.25) of \cite{25}):
\begin{equation}
\d_{\Lambda}{\cal Y}^i_{[\l_1\l_2\l_3]}(y) =
\cL_{\Lambda}{\cal Y}^i_{[\l_1\l_2\l_3]}(y),
\end{equation}
which satisfy the Lie algebra of the group $G$:
\begin{equation}
\left[\cL_{\Lambda},\cL_{\Sigma}\right]=
\cC^{\Pi}_{\ \Lambda\Sigma}\cL_{\Pi}.
\end{equation}
Moreover, the operators $\cL_{\Lambda}$ commute with the $SO(7)$
covariant derivative:
\begin{equation}\label{dercomm}
\cL_{\Lambda}D{\cal Y}^i=D\cL_{\Lambda}{\cal Y}^i\,,
\end{equation}
where $D$ is defined by
\begin{equation}\label{covder}
D=d+\cB^{\a\b}t_{\a\b},
\end{equation}
and $(t_{\a\b})^i_{\ j}$ are the generators of the $SO(7)$ irreducible
representation $[\l_1\l_2\l_3]$ of ${\cal Y}^i$ (e.g. $(t_{\a\b})^{\g\d}=-\d_{\a\b}^{\g\d}$
for the vector representation).
\par
An important thing to note is that $H=U(1)^{'}\times U(1)^{''}\times U(1)^{'''}$ is
necessarily a subgroup of $SO(7)$, whose natural embedding is given in terms of the
following embedding of the algebra $\IH$ into the adjoint representation of $SO(7)$:
\begin{eqnarray}
(T_H)^{\a}_{\ \b}=\cC^{\a}_{H\,\b},\nonumber\\
(T_{Z^{(1)}})^{\a}_{\ \b}=\frac{1}{\sqrt{3}}\left(
\begin{array}{cccc}
0&0&0&0\\
0&\e^{ij}&0&0\\
0&0&-\e^{i'j'}&0\\
0&0&0&0
\end{array}\right),\\
(T_{Z^{(2)}})^{\a}_{\ \b}=\frac{1}{\sqrt{3}}\left(
\begin{array}{cccc}
0&0&0&0\\
0&e^{ij}&0&0\\
0&0&\e^{i'j'}&0\\
0&0&0&-2\e^{i''j''}
\end{array}\right),\\
(T_{Z^{(3)}})^{\a}_{\ \b}=\ft{1}{\sqrt{3}}\left(
\begin{array}{cccc}
0&0&0&0\\
0&\e^{ij}&0&0\\
0&0&\e^{i'j'}&0\\
0&0&0&\e^{i''j''}
\end{array}\right).
\end{eqnarray}
This means that the $SO(7)$-indices of the various $n$-forms can be
split in different subsets, each one transforming into an irreducible
representation of $H$ and the $SO(7)$ irreducible representations $[\l_1\l_2\l_3]$ break into the direct
sum of $H$ irreducible representations.

\subsection{The zero-form}

Every harmonic is defined by its $H$ ([$Z^{(1)},Z^{(2)},Z^{(3)}$]) representation. The
only representation into which the $[0,0,0]$ (i.e. the scalar) of $SO(7)$ breaks under
$H$, is obviously the
$H$-scalar representation.\\
Solving the constraints (\ref{s1},\ref{s2},\ref{s3}) we see that
every harmonic is conveniently identified by the following labels:
\begin{equation}
\begin{tabular}{|c|c|c|}
  \hline
  $ \bf{J_{3}}$ & $ \bf{J_{3}'}$ & $ \bf{J_{3}''}$\\
  \hline \hline
  $k/2$ & $k/2$ & $k/2$ \\ \hline
\end{tabular}
\end{equation}
where $J_{3}$, $J_{3}'$, $J_{3}''$ denote the third components of the three isospin and
$k\in \mathbf{Z}$ is an integer number.\\
The Kaluza Klein mass operator for the zero-form ${\cal Y}$ is given by
\begin{equation}
\xbox^{[000]}{\cal Y} \equiv D_{\b}D^{\b}{\cal Y} = D^H_{\b}D^{H\b}{\cal Y}.
\end{equation}
 The computation of its
eigenvalues, on the $G$--representations, is immediate:
\begin{eqnarray}
&&\xbox^{[000]}{\cal Y} \equiv M_{(0)^3}{\cal Y} = H_0{\cal Y}\\
&&H_{0} \equiv 32\left[J(J+1)+J'(J'+1)+J''(J''+1)-\frac{k^2}{4}\right]\label{H}
\end{eqnarray}

\subsection{The one-form}

The decomposition under $H$ of the vector representation of
$SO(7)$ is the following:

\begin{equation}
[1,0,0] \to [0,0,0] \oplus [i, i, i]  \oplus [-i, -i, -i]
\oplus [i, -i, -i] \oplus [-i, i, i] \oplus [0, 2i, -i] \oplus [0 , -2i, i].
\end{equation}
where the three digits in the brackets [p,q,r] denote the eigenvalues of the generators
$Z^{(1)}$, $Z^{(2)}$ and $Z^{(3)}$.
 Every harmonic is then conveniently identified by the following
labels:
\begin{equation}
\begin{tabular}{|c|c|c|}

  \hline $\bf{J_{3}}$ & $\bf{J_{3}'}$ & $\bf{J_{3}''}$ \\
  \hline \hline
  $k/2$ & $k/2$ & $k/2$

   \\ \hline
  $k/2-1$ & $k/2$ & $k/2$ \\ \hline
  $k/2+1$ & $k/2$ & $k/2$ \\ \hline
  $k/2$ & $k/2-1$ & $k/2$ \\ \hline
  $k/2$ & $k/2+1$ & $k/2$ \\ \hline
  $k/2$ & $k/2$ & $k/2-1$ \\ \hline
  $k/2$ & $k/2$ & $k/2+1$ \\ \hline

\end{tabular}
 \end{equation}
 The Laplace Beltrami operator for the transverse
one-form field ${\cal Y}^{\a}$ is given by
\begin{equation}
\xbox^{[100]}{\cal Y}^{\a} \equiv M_{(1)(0)^2}{\cal Y}^{\a} = 2D_{\b}D^{[\b}{\cal Y}^{\a]} =
(D^{\b}D_{\b}+24){\cal Y}^{\a}\ ,
\end{equation}
where transversality of ${\cal Y}^{\a}$ means that $D_{\a}{\cal
Y}^{\a}=0$. From the decomposition $D_{\a}=D^{H}_{\a}+\IM_{\a}$
we obtain:
\begin{equation}
\xbox^{[100]}{\cal Y}^{\a} =
(D^{H\b}D^{H}_{\b}+24){\cal Y}^{\a}+\eta^{\g\d}\left(2(\IM_{\g})^{\a}_{\ \b}
D^H_{\d}+(\IM_{\g})^{\a}_{\ \e}(\IM_{\d})^{\e}_{\ \b}\right) {\cal Y}^{\b}\,.
\end{equation}
The matrix of this operator on the $AdS_4$ fields is given by

Column one to four:

\begin{footnotesize}
\begin{eqnarray}
\begin{array}{|c c  c c|}

H_0+48 & 8*\sqrt{2}(a+k/2+1) &  -8*\sqrt{2}(a-k/2+1) & 8*\sqrt{2}(b+k/2+1)  \cr
16\sqrt{2}(a-k/2) & H_0-16k & 0 & 0\cr

-16\sqrt{2}(a+k/2) & 0 & H_0+16k & 0 \cr

16\sqrt{2}(b-k/2) & 0 & 0 & H_0-16k \cr

-16\sqrt{2}(b+k/2) & 0 & 0 & 0 \cr

16\sqrt{2}(c-k/2) & 0 & 0 & 0 \cr

-16\sqrt{2}(c+k/2) & 0 & 0 & 0 \cr

\end{array}
\label{onematrix1}
\end{eqnarray}
\end{footnotesize}

Column five to seven:

\begin{footnotesize}
\begin{eqnarray}
\label{onematrix2}
\begin{array}{|c c c |}

   -8*\sqrt{2}(b-k/2+1) &  8*\sqrt{2}(c+k/2+1) &  -8*\sqrt{2}(c-k/2+1) \cr

 0 & 0 & 0 \cr

 0 & 0 & 0 \cr

 0 & 0 & 0 \cr

 H_0+16k & 0 & 0 \cr

 0 & H_0-16k& 0 \cr

 0 & 0 & H_0+16k\cr

\end{array}
\end{eqnarray}
\end{footnotesize}

This matrix has the following eigenvalues:
\begin{eqnarray}
\lambda_{1} &=& H_{0}
\nonumber \\
\lambda_{2} = \lambda_{3} &=& H_{0} - 16k
\nonumber \\
\lambda_{4} = \lambda_{5} &=& H_{0} + 16k
\nonumber \\
\lambda_{6} &=& H_{0} + 24 + 4\sqrt{H_{0}+36}
\nonumber \\
\lambda_{7} &=& H_{0}+24 -4\sqrt{H_{0} + 36}
\nonumber \\
\label{eigenAoneform}
\end{eqnarray}
in terms of $H_0$ defined in (\ref{H})

Our actual operator (\ref{onematrix1}, \ref{onematrix2}) contains the eigenvalues of
$M_{(0)^3}$, which are \emph{longitudinal} (hence \emph{non-physical}) for the one--form.

The eigenvalue $\lambda_1$ in \eqn{eigenAoneform} is the longitudinal one,
equal to the zero--form eigenvalue $H_0$.

\subsection{The two-form mass operator}

Under the action of $H=U(1)'\times U(1)''\times U(1)^{'''}$ the 21 components of the
$SO(7)$ two-form transform into the completely reducible representation:
\begin{eqnarray}
[1,1,0] & \to & [0,2i,2i] \oplus [0,-2i,-2i] \oplus [i,-i,2i]
\oplus [-i, i,-2i] \oplus [-i,-i,2i] \oplus\nonumber\\
 && \oplus [i,i,-2i] \oplus [2i,0,0] \oplus [-2i,0,0]
\oplus [i,3i,0] \nonumber\\
 && \oplus [-i,-3i,0] \oplus [-i,3i,0] \oplus [i,-3i,0]
\oplus 3[0,0,0], \nonumber\\ && \oplus   [i, i, i]  \oplus [-i, -i, -i]
\oplus [i, -i, -i] \oplus [-i, i, i] \oplus[0, 2i, -i] \oplus [0 , -2i, i].
\end{eqnarray}
Every harmonic is then conveniently identified by the following
labels:
\begin{equation}
\begin{tabular}{|c|c|c|}
  \hline
  $\bf{J_{3}}$ & $\bf{J_{3}'}$ & $\bf{J_{3}''}$\\
  \hline\hline
  $k/2+1$ & $k/2+1$ & $k/2$  \\ \hline
  $k/2-1$  & $k/2-1$  & $k/2$  \\ \hline
  $k/2+1$  & $k/2$  & $k/2+1$  \\ \hline
  $k/2-1$  & $k/2$  & $k/2-1$  \\ \hline
  $k/2$  & $k/2+1$  & $k/2+1$  \\ \hline
  $k/2$  & $k/2-1$  & $k/2-1$  \\ \hline
  $k/2+1$  & $k/2-1$  & $k/2$  \\ \hline
  $k/2-1$  & $k/2+1$  & $k/2$  \\ \hline
  $k/2+1$  & $k/2$  & $k/2-1$  \\ \hline
  $k/2-1$  & $k/2$  & $k/2+1$  \\ \hline
  $k/2$  & $k/2+1$  & $k/2-1$  \\ \hline
  $k/2$  & $k/2-1$  & $k/2+1$  \\ \hline
  $k/2$  & $k/2$  & $k/2$  \\ \hline
  $k/2$  & $k/2$  & $k/2$  \\ \hline
  $k/2$  & $k/2$  & $k/2$  \\ \hline
  $k/2+1$  & $k/2$  & $k/2$  \\ \hline
  $k/2-1$  & $k/2$  & $k/2$  \\ \hline
  $k/2$  & $k/2+1$  & $k/2$  \\ \hline
  $k/2$  & $k/2-1$  & $k/2$  \\ \hline
  $k/2$  & $k/2$  & $k/2+1$  \\ \hline
  $k/2$  & $k/2$  & $k/2-1$  \\ \hline
\end{tabular}
\end{equation}
 The Laplace Beltrami operator for the transverse two-form
field ${\cal Y}^{\a\b}$, is given by
\begin{equation}
\xbox^{[110]}{\cal Y}^{[\a\b]} \equiv M_{(1)^2(0)}{\cal Y}^{[\a\b]} =
3D_{\g}D^{[\g}{\cal Y}^{\a\b]} = (D^{\g}D_{\g}+48){\cal Y}^{[\a\b]}-
4\cR^{[\a\ \b]}_{\ [\g\ \d]}{\cal Y}^{[\g\d]}
\end{equation}
From the decomposition $D_{\a}{\cal Y}^{\b\g}=D^H_{\a}{\cal Y}^{\b\g}
+(\IM_{\a})^{\b}_{\ \d}{\cal Y}^{\d\g}+(\IM_{\a})^{\g}_{\ \d}{\cal Y}^{\b\d}$
we obtain:
\begin{eqnarray}
\xbox^{[110]}{\cal Y}^{[\a\b]} = \left\{48\,\d^{[\a\ \b]}_{\ [\g\ \d]}-
4\cR^{[\a\ \b]}_{\ [\g\ \d]} + \right. & \nonumber\\
 & \nonumber\\
2\eta^{\mu\nu}(\IM_{\mu})^{[\a}_{\ \,[\g}(\IM_{\nu})^{\b]}_{\ \,\d]}+
2\eta^{\mu\nu}(\IM_{\mu}\IM_{\nu})_{\ [\g}^{[\a}\d^{\b]}_{\ \d]}\, + & \!\!\!\left.
4\eta^{\mu\nu}(\IM_{\mu})_{\ [\g}^{[\a}\d^{\b]}_{\ \d]}D^H_{\nu}
\right\} {\cal Y}^{[\g\d]}\,.
\end{eqnarray}
This operators acts on
$AdS_4$ fields as the following $21\times 21$ matrix:\\
Column one to five:
\begin{footnotesize}
\begin{eqnarray*}
\begin{array}{|c c c c c|}

H_0-32k & 0 & 0 & 0 & 0 \cr

0 & H_0+32k & 0 & 0 & 0 \cr

0 & 0 &H_0-32k  & 0 & 0 \cr

0 & 0 & 0 & H_0+32k & 0 \cr

0 & 0 & 0 & 0 & H_0-32k \cr

0 & 0 & 0 & 0 & 0 \cr

0 & 0 & 0 & 0 & 0 \cr

0 & 0 & 0 & 0 & 0 \cr

0 & 0 & 0 & 0 & 0 \cr

0 & 0 & 0 & 0 & 0 \cr

0 & 0 & 0 & 0 & 0 \cr

0 & 0 & 0 & 0 & 0 \cr

0 & 0 & 0 & 0 & 0 \cr

0 & 0 & 0 & 0 & 0 \cr

0 & 0 & 0 & 0 & 0 \cr

2\sqrt{32}(b+\frac{k}{2}+1) & 0 & 2\sqrt{32}(c+\frac{k}{2}+1) & 0 & 0 \cr

0 & -2\sqrt{32}(b-\frac{k}{2}+1) & 0 & -2\sqrt{32}(c-\frac{k}{2}+1) & 0 \cr

 -2\sqrt{32}(a+\frac{k}{2}+1) & 0 & 0 & 0 &  2\sqrt{32}(c+\frac{k}{2}+1) \cr

0 & 2\sqrt{32}(a-\frac{k}{2}+1) & 0 & 0 & 0 \cr

0 & 0 &  -2\sqrt{32}(a+\frac{k}{2}+1) & 0 &  -2\sqrt{32}(b+\frac{k}{2}+1) \cr

0 & 0 & 0 &  2\sqrt{32}(a-\frac{k}{2}+1) & 0 \cr

\end{array}
\end{eqnarray*}
\end{footnotesize}

Column six to ten:
\begin{footnotesize}
\begin{eqnarray*}
\begin{array}{|c c c c c|}

0 & 0 & 0 & 0 & 0 \cr

0 & 0 & 0 & 0 & 0 \cr

0 & 0 & 0 & 0 & 0 \cr

0 & 0 & 0 & 0 & 0 \cr

0 & 0 & 0 & 0 & 0 \cr

H_0+32k & 0 & 0 & 0 & 0 \cr

0 & H_0 & 0 & 0 & 0 \cr

0 & 0 & H_0 & 0 & 0 \cr

0 & 0 & 0 & H_0 & 0 \cr

0 & 0 & 0 & 0 & H_0 \cr

0 & 0 & 0 & 0 & 0 \cr

0 & 0 & 0 & 0 & 0 \cr

0 & 0 & 0 & 0 & 0 \cr

0 & 0 & 0 & 0 & 0 \cr

0 & 0 & 0 & 0 & 0 \cr

0 &  -2\sqrt{32}(b-\frac{k}{2}+1) & 0 &  -2\sqrt{32}(c-\frac{k}{2}+1) & 0 \cr

0 & 0 &  2\sqrt{32}(b+\frac{k}{2}+1) & 0 &  2\sqrt{32}(c+\frac{k}{2}+1) \cr

0 & 0 &  2\sqrt{32}(a-\frac{k}{2}+1) & 0 & 0 \cr

 -2\sqrt{32}(c-\frac{k}{2}+1) &  -2\sqrt{32}(a+\frac{k}{2}+1) & 0 & 0 & 0 \cr

0 & 0 & 0 & 0 &  2\sqrt{32}(a-\frac{k}{2}+1) \cr

 2\sqrt{32}(b-\frac{k}{2}+1) & 0 & 0 &  -2\sqrt{32}(a+\frac{k}{2}+1) & 0 \cr

\end{array}
\end{eqnarray*}
\end{footnotesize}
Column eleven to sixteen:
\begin{footnotesize}
\begin{eqnarray*}
\begin{array}{|c c c c c c|}

0 & 0 & 0 & 0 & 0 & 4\sqrt{32}(b-\frac{k}{2})  \cr

0 & 0 & 0 & 0 & 0 & 0 \cr

0 & 0 & 0 & 0 & 0 &  4\sqrt{32}(c-\frac{k}{2}) \cr

0 & 0 & 0 & 0 & 0  & 0 \cr

0 & 0 & 0 & 0 & 0  & 0\cr

0 & 0 & 0 & 0 & 0  & 0\cr

0 & 0 & 0 & 0 & 0 &  -4\sqrt{32}(b+\frac{k}{2})\cr

0 & 0 & 0 & 0 & 0  & 0\cr

0 & 0 & 0 & 0 & 0 &  -4\sqrt{32}(c+\frac{k}{2})\cr

0 & 0 & 0 & 0 & 0  & 0\cr

H_0 & 0 & 0 & 0 & 0  & 0\cr

0 & H_0 & 0 & 0 & 0  & 0\cr

0 & 0 & H_0+16 & 16 & 16 & 2{\mathrm i}\sqrt{32}(a+\frac{k}{2}+1) \cr

0 & 0 & 16 & H_0+16 & 16 & 0 \cr

0 & 0 & 16 & 16 & H_0+16  & 0\cr

0 & 0 & -4{\mathrm i}\sqrt{32}(a-\frac{k}{2}) & 0 & 0 & H_0+32-16k\cr

0 & 0 & -4{\mathrm i}\sqrt{32}(a+\frac{k}{2}) & 0 & 0  & 0\cr

-2\sqrt{32}(c-\frac{k}{2}+1) & 0 & 0 & -4{\mathrm i}\sqrt{32}(b-\frac{k}{2}) & 0  & 0\cr

0 & 2\sqrt{32}(c+\frac{k}{2}+1) & 0 & -4{\mathrm i}\sqrt{32}(b+\frac{k}{2}) & 0  & 0\cr

0 & 2\sqrt{32}(b-\frac{k}{2}+1) & 0 & 0 & -4{\mathrm i}\sqrt{32}(c-\frac{k}{2})  & 0\cr

-2\sqrt{32}(b+\frac{k}{2}+1) & 0 & 0 & 0 & -4{\mathrm i}\sqrt{32}(c+\frac{k}{2})  & 0\cr

\end{array}
\end{eqnarray*}
\end{footnotesize}

Column seventeen to twentyone:
\begin{footnotesize}
\begin{eqnarray*}
\begin{array}{| c c  c c c |}

0 & -4\sqrt{32}(a-\frac{k}{2}) & 0 & 0 & 0 \cr

 -4\sqrt{32}(b+\frac{k}{2}) & 0 &  4\sqrt{32}(a+\frac{k}{2}) & 0 & 0 \cr

0 & 0 & 0 &  -4\sqrt{32}(a-\frac{k}{2}) & 0 \cr

 -4\sqrt{32}(c+\frac{k}{2}) & 0 & 0 & 0 &  4\sqrt{32}(a+\frac{k}{2}) \cr

0 &  4\sqrt{32}(c-\frac{k}{2}) & 0 &  -4\sqrt{32}(b-\frac{k}{2}) & 0 \cr

0 & 0 &  -4\sqrt{32}(c+\frac{k}{2}) & 0 &  4\sqrt{32}(b+\frac{k}{2}) \cr

0 & 0 & -4\sqrt{32}(a-\frac{k}{2}) & 0 & 0 \cr

 4\sqrt{32}(b-\frac{k}{2}) &  4\sqrt{32}(a+\frac{k}{2}) & 0 & 0 & 0 \cr

0 & 0 & 0 & 0 &  -4\sqrt{32}(a-\frac{k}{2}) \cr

 4\sqrt{32}(c-\frac{k}{2}) & 0 & 0 &  4\sqrt{32}(a+\frac{k}{2}) & 0 \cr

0 &  -4\sqrt{32}(c+\frac{k}{2}) & 0 & 0 &  -4\sqrt{32}(b-\frac{k}{2}) \cr

0 & 0 &  4\sqrt{32}(c-\frac{k}{2}) &  4\sqrt{32}(b+\frac{k}{2}) & 0 \cr

2{\mathrm i} \sqrt{32}(a-\frac{k}{2}+1) & 0 & 0 & 0 & 0 \cr

0 & 2{\mathrm i} \sqrt{32}(b+\frac{k}{2}+1) & 2{\mathrm i} \sqrt{32}(b-\frac{k}{2}+1) & 0
& 0 \cr

0 & 0 & 0 & 2{\mathrm i} \sqrt{32}(c+\frac{k}{2}+1) & 2{\mathrm i}
\sqrt{32}(c-\frac{k}{2}+1) \cr

0 & 0 & 0 & 0 & 0 \cr

H_0+32+16k & 0 & 0 & 0 & 0 \cr

0 & H_0+32-16k & 0 & 0 & 0 \cr

0 & 0 & H_0+32+16k & 0 & 0 \cr

0 & 0 & 0 & H_0+32-16k & 0 \cr

0 & 0 & 0 & 0 & H_0+32+16k \cr

\end{array}
\end{eqnarray*}
\end{footnotesize}

This matrix has the following eigenvalues:
\begin{eqnarray}
\lambda_1 = \lambda_2 \lambda_3 &=& H_0
\nonumber \\
\lambda_4 = \lambda_5 &=& H_{0} - 16k
\nonumber \\
\lambda_6 = \lambda_7 &=& H_{0} + 16k
\nonumber \\
\lambda_8 &=& H_{0} + 24 + 4\sqrt{H_{0}+36}
\nonumber \\
\lambda_9 &=& H_{0}+24 -4\sqrt{H_{0} + 36}
\nonumber \\
\lambda_{10} = \lambda_{11} &=& H_0+32
\nonumber
\\
\lambda_{12} &=& H_0-32k
\nonumber \\
\lambda_{13} &=& H_0+32k
\nonumber \\
\lambda_{14}=\lambda_{15} &=& H_0+16k+16 +4\sqrt{H_0+16k+16}
\nonumber\\
\lambda_{16}=\lambda_{17} &=& H_0+16k+16 -4\sqrt{H_0+16k+16}
\nonumber\\
\lambda_{18}=\lambda_{19} &=& H_0-16k+16 -4\sqrt{H_0-16k+16}
\nonumber\\
\lambda_{20}=\lambda_{21} &=& H_0-16k+16 +4\sqrt{H_0-16k+16} \label{eigenAtwoform}
\end{eqnarray}

The eigenvalues $\lambda_1,\lambda_4,\lambda_5,\lambda_6\lambda_7,\lambda_8,\lambda_9$
are those associated with longitudinal two forms that are non physical since they can be
removed by gauge transformations. The remaining eigenvalues, associated with transverse
two forms, provide the spectrum of physical states.

\subsection{The three-form mass operator}

The $H$ decomposition of the three form into $H$ irreducible fragments is the following:

\begin{eqnarray}
[1,1,1] & \to & [0,2i,2i] \oplus 3[0,0,0] \oplus [0,2i,2i]
\oplus [2i, 0,0] \oplus [i,-i,2i] \oplus\nonumber\\
 && \oplus [i,3i,0] \oplus [-2i,0,0] \oplus [0,-2i,-2i]
\oplus [-i,-3i,0]
  \oplus [-i,i,-2i] \oplus [-i,-i,2i] \nonumber\\ &&\oplus [-i,3i,0]
\oplus 3[i,-3i,0],  \oplus   [i, i, -2i]  \oplus [0, 0, 3i]
\oplus [0, 4i, i] \nonumber\\ &&\oplus [2i, -2i, i] \oplus[2i, 2i, -i] \oplus [-2i , -2i, i]
\oplus 2[i,i,i] \oplus 2[-i,-i,-i] \nonumber\\ &&
\oplus 2[-i,i,i] \oplus 2[i,-i,-i] \oplus 2[0,-2i,i] \oplus 2[0, 2i, -i] \oplus [-2i,2i,-i] \nonumber\\ && \oplus [0,-4i,-i] \oplus [0,0,-3i]
\end{eqnarray}
Every harmonic is then conveniently identified by the following
labels:
\begin{equation}
\begin{tabular}{|c|c|c|}
  \hline
  $\bf{J_{3}}$& $\bf{J_{3}'}$ & $\bf{J_{3}''}$ \\
  \hline\hline
  $k/2$ & $k/2$ & $k/2$  \\ \hline
  $k/2+1$  & $k/2+1$  & $k/2$  \\ \hline
  $k/2+1$  & $k/2-1$  & $k/2$  \\ \hline
  $k/2+1$  & $k/2$  & $k/2+1$  \\ \hline
  $k/2+1$  & $k/2$  & $k/2-1$  \\ \hline
  $k/2-1$  & $k/2+1$  & $k/2$  \\ \hline
  $k/2-1$  & $k/2-1$  & $k/2$  \\ \hline
  $k/2-1$  & $k/2$  & $k/2+1$  \\ \hline
  $k/2-1$  & $k/2$  & $k/2-1$  \\ \hline
  $k/2$ & $k/2$ & $k/2$ \\ \hline
  $k/2$ & $k/2+1$ & $k/2+1$ \\ \hline
  $k/2$ & $k/2+1$ & $k/2-1$ \\ \hline
  $k/2$ & $k/2-1$ & $k/2+1$ \\ \hline
  $k/2$ & $k/2-1$ & $k/2-1$ \\ \hline
  $k/2$ & $k/2$ & $k/2$ \\ \hline
  $k/2$ & $k/2+1$ & $k/2$ \\ \hline
  $k/2$ & $k/2-1$ & $k/2$ \\ \hline
  $k/2$ & $k/2$ & $k/2+1$ \\ \hline
  $k/2$ & $k/2$ & $k/2-1$ \\ \hline
  $k/2+1$ & $k/2$ & $k/2$ \\ \hline
  $k/2+1$ & $k/2+1$ & $k/2+1$ \\ \hline
  $k/2+1$ & $k/2+1$ & $k/2-1$ \\ \hline
  $k/2+1$ & $k/2-1$ & $k/2+1$ \\ \hline
  $k/2+1$ & $k/2-1$ & $k/2-1$ \\ \hline
  $k/2+1$ & $k/2$ & $k/2$ \\ \hline
  $k/2-1$ & $k/2$ & $k/2$ \\ \hline
  $k/2-1$ & $k/2+1$ & $k/2+1$ \\ \hline
  $k/2-1$ & $k/2+1$ & $k/2-1$ \\ \hline
  $k/2-1$ & $k/2-1$ & $k/2+1$ \\ \hline
  $k/2-1$ & $k/2-1$ & $k/2-1$ \\ \hline
  $k/2-1$ & $k/2$ & $k/2$ \\ \hline
  $k/2$ & $k/2$ & $k/2+1$ \\ \hline
  $k/2$ & $k/2$ & $k/2-1$ \\ \hline
  $k/2$ & $k/2+1$ & $k/2$ \\ \hline
  $k/2$ & $k/2-1$ & $k/2$ \\ \hline

\end{tabular}
\end{equation}
The Laplace Beltrami operator for the transverse three-form
${\cal Y}^{[\a\b\g]}$, is a first-order differential operator, given by
\begin{eqnarray}
\xbox^{[111]}{\cal Y}^{[\a\b\g]} \equiv M_{(1)^3}{\cal Y}^{[\a\b\g]} =
\ft{1}{24}\e^{\a\b\g\d}_{\ \ \ \ \mu\nu\r}
D_{\d}{\cal Y}^{\mu\nu\r} =\hspace{6 cm}\nonumber\\
=\ft{1}{24}\e^{\a\b\g\d}_{\ \ \ \ \ \mu\nu\r}\left[
D^H_{\d}{\cal Y}^{\mu\nu\r}+(\IM_{\d})^{\mu}_{\ \s}{\cal Y}^{\s\nu\r}+
(\IM_{\d})^{\nu}_{\ \s}{\cal Y}^{\mu\s\r}+
(\IM_{\d})^{\r}_{\ \s}{\cal Y}^{\mu\nu\s}\right].
\end{eqnarray}
This operator acts on the $AdS_4$ fields as a
$35\times 35$ matrix:\\
Column one to six:
\begin{scriptsize}
\begin{equation}
\begin{array}{| c c c c c c|}

  0 & 0 &  0 & 0 & 0 & 0 \cr

  0 & 4 & 0 & 0 & 0 & 0  \cr

  0 & 0 & -4 & 0 & 0 & 0  \cr

  0 & 0  &0  & 4 & 0 & 0  \cr

  0 & 0  & 0 & 0 & -4 & 0   \cr

  0 & 0  & 0 & 0 & 0 & -4 \cr

  0 & 0  & 0 & 0 & 0 & 0 \cr

  0 & 0  & 0 & 0 & 0 & 0 \cr

  0 & 0  & 0 & 0 & 0 & 0  \cr

  4 & 0  & 0 & 0 & 0 & 0 \cr

  0 & 0  & 0 & 0 & 0 & 0\cr

  0 & 0  & 0 & 0 & 0 & 0 \cr

  0 & 0  & 0 & 0 & 0 & 0 \cr

  0 & 0  & 0 & 0 & 0 & 0 \cr

  4 & 0  & 0 & 0 & 0 & 0 \cr

  0 & 0  & 0 & 0 & 0 & 0 \cr

  0 & 0  & 0 & 0 & 0 & 0 \cr

  0 & 0  & 0 & 0 & 0 & 0  \cr

  0 & 0  & 0 & 0 & 0 & 0 \cr

  0 & 0  & 0 & -\frac{1}{2}\sqrt{32}(c+\frac{k}{2}+1) &\frac{1}{2}\sqrt{32}(c-\frac{k}{2}+1)  & 0 \cr

  0 & -\sqrt{32}(c-\frac{k}{2})  & 0 & \sqrt{32}(b-\frac{k}{2}) & 0 & 0 \cr

  0 & \sqrt{32}(c+\frac{k}{2})  & 0 & 0 & -\sqrt{32}(b-\frac{k}{2}) & 0 \cr

  0 & 0  & \sqrt{32}(c-\frac{k}{2}) & -\sqrt{32}(b+\frac{k}{2}) & 0 & 0 \cr

  0 & 0  & - \sqrt{32}(c+\frac{k}{2} ) & 0 & \sqrt{32}(b+\frac{k}{2})  & 0 \cr

  0 & -\frac{1}{2} \sqrt{32}(b+\frac{k}{2}+1)  & \frac{1}{2} \sqrt{32}(b-\frac{k}{2}+1) & 0 & 0 & 0 \cr

  0 & 0  & 0 & 0 & 0 & 0 \cr

  0 & 0  & 0 & 0 & 0 &  \sqrt{32}(c-\frac{k}{2}) \cr

  0 & 0  & 0 & 0 & 0 & - \sqrt{32}(c+\frac{k}{2}) \cr

  0 & 0  & 0 & 0 & 0 & 0 \cr

  0 & 0  & 0 & 0 & 0 & 0 \cr

  0 & 0  & 0 & 0 & 0 & \frac{1}{2} \sqrt{32}(b+\frac{k}{2}+1) \cr

  -{\mathrm i} \sqrt{32}(c-\frac{k}{2}) & 0  & 0 & \frac{1}{2} \sqrt{32}(a+\frac{k}{2}+1) & 0 & 0
\cr

 {\mathrm i} \sqrt{32}(c+\frac{k}{2})& 0 & 0  & 0 & -\frac{1}{2} \sqrt{32}(a+\frac{k}{2}+1) & 0  \cr

   -{\mathrm i} \sqrt{32}(b-\frac{k}{2}) & \frac{1}{2} \sqrt{32}(a+\frac{k}{2}+1)  & 0 & 0 & 0 &
-\frac{1}{2} \sqrt{32}(a-\frac{k}{2}+1)\cr

  {\mathrm i} \sqrt{32}(b+\frac{k}{2}) & 0  &  -\frac{1}{2} \sqrt{32}(a+\frac{k}{2}+1) & 0 & 0 & 0  \cr
\end{array}
\end{equation}
\end{scriptsize}
\newpage
Column seven to eleven:
\begin{footnotesize}

\begin{eqnarray}
\begin{array}{|c c c c c|}

   0 & 0 & 0 & 4 & 0 \cr

   0 & 0 & 0 & 0 & 0 \cr

   0 & 0 & 0 & 0 & 0  \cr

   0 & 0 & 0 & 0 & 0  \cr

  0 & 0  & 0 & 0 & 0 \cr

   0 & 0 & 0 & 0 & 0  \cr

   4 & 0 & 0 & 0 & 0  \cr

   0 & -4 & 0 & 0 & 0  \cr

   0 & 0  & 4 & 0 & 0 \cr

   0 & 0 & 0 & 0 & 0  \cr

   0 & 0 & 0 & 0 & 4  \cr

   0 & 0 & 0 & 0 & 0  \cr

   0 & 0 & 0 & 0 & 0   \cr

   0 & 0 & 0 & 0 & 0  \cr

   0 & 0  & 0 & 4 & 0\cr

   0 & 0  & 0 & 0 & -\frac{1}{2}\sqrt{32}(c+\frac{k}{2}+1)\cr

   0 & 0  & 0 & 0 & 0 \cr

   0 & 0  & 0 &-{\mathrm i}\sqrt{32}(c-\frac{k}{2}) & \frac{1}{2}\sqrt{32}(b+\frac{k}{2}+1)  \cr

   0 & 0  & 0 & {\mathrm i}\sqrt{32}(c+\frac{k}{2}) & 0 \cr

  0 & 0  & 0 & 0 & 0\cr

   0 & 0  & 0 & 0 & -\sqrt{32}(a-\frac{k}{2}) \cr

   0 & 0  & 0 & 0 & 0 \cr

   0 & 0  & 0 & 0 & 0\cr

   0 & 0  & 0 & 0 & 0 \cr

   0 & 0  & 0 & -{\mathrm i}\sqrt{32}(a-\frac{k}{2}) & 0 \cr

   0 & \frac{1}{2} \sqrt{32}(c+\frac{k}{2}+1) & -\frac{1}{2} \sqrt{32}(c-\frac{k}{2}+1) & 0 & 0 \cr

   0 & - \sqrt{32}(b-\frac{k}{2}) & 0 & 0 & \sqrt{32}(a+\frac{k}{2}) \cr

   0 & 0  & \sqrt{32}(b-\frac{k}{2})  & 0 & 0 \cr

   - \sqrt{32}(c-\frac{k}{2}) &  \sqrt{32}(b+\frac{k}{2}) & 0 & 0 & 0   \cr

	\sqrt{32}(c+\frac{k}{2})& 0  &  -\sqrt{32}(b+\frac{k}{2}) & 0 & 0 \cr

  -\frac{1}{2} \sqrt{32}(b-\frac{k}{2}+1) & 0  & 0 &{\mathrm i}\sqrt{32}(a+\frac{k}{2})  & 0 \cr

  0 & - \frac{1}{2} \sqrt{32}(a-\frac{k}{2}+1)  & 0 & 0 & 0 \cr

   0 & 0  &   \frac{1}{2} \sqrt{32}(a-\frac{k}{2}+1)   & 0 & 0 \cr

	0 & 0  & 0 & 0 & 0 \cr

   \frac{1}{2} \sqrt{32}(a-\frac{k}{2}+1) & 0  & 0 & 0 & 0 \cr

  \end{array}
\end{eqnarray}
\end{footnotesize}
Column twelve to sixteen:
\begin{footnotesize}
\begin{eqnarray}
\begin{array}{|c c c c c|}

0 & 0 & 0 & 4 & 0 \cr

0 & 0 & 0 & 0 & 0 \cr

0 & 0 & 0 & 0 & 0 \cr

0 & 0 & 0 & 0 & 0 \cr

0 & 0 & 0 & 0 & 0 \cr

0 & 0 & 0 & 0 & 0 \cr

0 & 0 & 0 & 0 & 0 \cr

0 & 0 & 0 & 0 & 0 \cr

0 & 0 & 0 & 0 & 0 \cr

0 & 0 & 0 & 4 & 0 \cr

0 & 0 & 0 & 0 & -\sqrt{32}(c-\frac{k}{2}) \cr

-4 & 0 & 0 & 0 & \sqrt{32}(c+\frac{k}{2}) \cr

0 & -4 & 0 & 0 & 0 \cr

0 & 0 & 4 & 0 & 0 \cr

0 & 0 & 0 & 0 & \frac{1}{2}{\mathrm i}\sqrt{32}(b+\frac{k}{2}+1) \cr

\frac{1}{2}\sqrt{32}(c-\frac{k}{2}+1)  & 0 & 0 &-{\mathrm i}\sqrt{32}(b-\frac{k}{2}) & 0
\cr

0 &\frac{1}{2}\sqrt{32}(c+\frac{k}{2}+1)  & -\frac{1}{2}\sqrt{32}(c-\frac{k}{2}+1)
&{\mathrm i}\sqrt{32}(b+\frac{k}{2})  & 0 \cr

0 & -\frac{1}{2}\sqrt{32}(b-\frac{k}{2}+1) & 0 & 0 & 0 \cr

-\frac{1}{2}\sqrt{32}(b+\frac{k}{2}+1) & 0 & \frac{1}{2}\sqrt{32}(b-\frac{k}{2}+1) & 0 &
0 \cr

0 & 0 & 0 & -{\mathrm i}\sqrt{32}(a-\frac{k}{2})& 0 \cr

0 & 0 & 0 & 0 & 0 \cr

\sqrt{32}(a-\frac{k}{2}) & 0 & 0 & 0 & 0 \cr

0 & \sqrt{32}(a-\frac{k}{2}) & 0 & 0 & 0 \cr

0 & 0 &-\sqrt{32}(a-\frac{k}{2})  & 0 & 0 \cr

0 & 0 & 0 & 0 & 0 \cr

0 & 0 & 0 &{\mathrm i}\sqrt{32}(a+\frac{k}{2}) & 0 \cr

0 & 0 & 0 & 0 & 0 \cr

-{\mathrm i}\sqrt{32}(a+\frac{k}{2}) & 0 & 0 & 0 & 0 \cr

0 & -\sqrt{32}(a+\frac{k}{2}) & 0 & 0 & 0 \cr

0 & 0 & \sqrt{32}(a+\frac{k}{2}) & 0 & 0 \cr

0 & 0 & 0 & 0 & 0 \cr

0 & 0 & 0 & 0 & 0 \cr

0 & 0 & 0 & 0 & 0 \cr

0 & 0 & 0 & 0 & -4k \cr

0 & 0 & 0 & 0 & 0 \cr

\end{array}
\end{eqnarray}
\end{footnotesize}

\newpage
 Column seventeen to twentyone:

\begin{footnotesize}
\begin{eqnarray}
\begin{array}{|c c c c c|}

0 & 0 & 0 & 0 & 0 \cr

0 & 0 & 0 & 0 & -\frac{1}{2}\sqrt{32}(c+\frac{k}{2}+1) \cr

0 & 0 & 0 & 0 & 0 \cr

0 & 0 & 0 & -\sqrt{32}(c-\frac{k}{2}) & \frac{1}{2}\sqrt{32}(b+\frac{k}{2}+1) \cr

0 & 0 & 0 & \sqrt{32}(c+\frac{k}{2}) & 0 \cr

0 & 0 & 0 & 0 & 0 \cr

0 & 0 & 0 & 0 & 0 \cr

0 & 0 & 0 & 0 & 0 \cr

0 & 0 & 0 & 0 & 0 \cr

0 & \frac{1}{2}{\mathrm i}\sqrt{32}(c+\frac{k}{2}+1) & -\frac{1}{2}{\mathrm i}
\sqrt{32}(c-\frac{k}{2}+1) & 0 & 0 \cr

0 & \sqrt{32}(b-\frac{k}{2})& 0 & 0 & -\frac{1}{2}\sqrt{32}(a+\frac{k}{2}+1) \cr

0 & 0 & -\sqrt{32}(b-\frac{k}{2}) & 0 & 0 \cr

\sqrt{32}(c-\frac{k}{2}) & -\sqrt{32}(b+\frac{k}{2}) & 0 & 0 & 0 \cr

-\sqrt{32}(c+\frac{k}{2}) & 0 & 0 & 0 & 0 \cr

-\frac{1}{2}{\mathrm i}\sqrt{32}(b-\frac{k}{2}+1)& 0 & 0 & \frac{1}{2}{\mathrm
i}\sqrt{32}(a+\frac{k}{2}+1)   & 0 \cr

0 & 0 & 0 & 0 & 0 \cr

0 & 0 & 0 & 0 & 0 \cr

0 & 0 & 0 & 0 & 0 \cr

0 & 0 & 0 & 0 & 0 \cr

0 & 0 & 0 & 0 & 0 \cr

0 & 0 & 0 & 0 & -4k \cr

0 & 0 & 0 & 0 & 0 \cr

0 & 0 & 0 & 0 & 0 \cr

0 & 0 & 0 & 0 & 0 \cr

0 & 0 & 0 & -4k & 0 \cr

0 & 0 & 0 & 0 & 0 \cr

0 & 0 & 0 & 0 & 0 \cr

0 & 0 & 0 & 0 & 0 \cr

0 & 0 & 0 & 0 & 0 \cr

0 & 0 & 0 & 0 & 0 \cr

0 & 0 & 0 & 0 & 0 \cr

0 & -4k & 0 & 0 & 0 \cr

0 & 0 & 4k & 0 & 0 \cr

0 & 0 & 0 & 0 & 0 \cr

4k & 0 & 0 & 0 & 0 \cr

\end{array}
\end{eqnarray}
\end{footnotesize}

\newpage
Column twentytwo to twentysix:

\begin{footnotesize}
\begin{eqnarray}
\begin{array}{|c c c c c|}

0 & 0 & 0 & 0 & 0 \cr

\frac{1}{2} \sqrt{32}(c-\frac{k}{2}+1)  & 0 & 0 & -\sqrt{32}(b-\frac{k}{2}) & 0 \cr

0 & \frac{1}{2} \sqrt{32}(c+\frac{k}{2}+1)  &-\frac{1}{2} \sqrt{32}(c-\frac{k}{2}+1)   &
\sqrt{32}(b+\frac{k}{2}) & 0 \cr

0 &-\frac{1}{2} \sqrt{32}(b-\frac{k}{2}+1)   & 0 & 0 & 0 \cr

-\frac{1}{2} \sqrt{32}(b+\frac{k}{2}+1)  & 0 & \frac{1}{2} \sqrt{32}(b-\frac{k}{2}+1)  &
0 & 0 \cr

0 & 0 & 0 & 0 & 0 \cr

0 & 0 & 0 & 0 & 0 \cr

0 & 0 & 0 & 0 & \sqrt{32}(c-\frac{k}{2}) \cr

0 & 0 & 0 & 0 & -\sqrt{32}(c+\frac{k}{2})  \cr

0 & 0 & 0 & \frac{1}{2}{\mathrm i}\sqrt{32}(a+\frac{k}{2}+1) & 0 \cr

0 & 0 & 0 & 0 & 0 \cr

 \frac{1}{2}\sqrt{32}(a+\frac{k}{2}+1) & 0 & 0 & 0 & 0 \cr

0 &  \frac{1}{2}\sqrt{32}(a+\frac{k}{2}+1) & 0 & 0 & 0 \cr

0 & 0 &  -\frac{1}{2}\sqrt{32}(a+\frac{k}{2}+1) & 0 & 0 \cr

0 & 0 & 0 & 0 & -\frac{1}{2}{\mathrm i}\sqrt{32}(a-\frac{k}{2}+1) \cr

0 & 0 & 0 & 0 & 0 \cr

0 & 0 & 0 & 0 & 0 \cr

0 & 0 & 0 & 0 & 0 \cr

0 & 0 & 0 & 0 & 0 \cr

0 & 0 & 0 & -4k & 0 \cr

0 & 0 & 0 & 0 & 0 \cr

4k & 0 & 0 & 0 & 0 \cr

0 & 4k & 0 & 0 & 0 \cr

0 & 0 & -4k & 0 & 0 \cr

0 & 0 & 0 & 0 & 0 \cr

0 & 0 & 0 & 0 & 0 \cr

0 & 0 & 0 & 0 & 0 \cr

0 & 0 & 0 & 0 & 0 \cr

0 & 0 & 0 & 0 & 0 \cr

0 & 0 & 0 & 0 & 0 \cr

0 & 0 & 0 & 0 & 4k \cr

0 & 0 & 0 & 0 & 0 \cr

0 & 0 & 0 & 0 & 0 \cr

0 & 0 & 0 & 0 & 0 \cr

0 & 0 & 0 & 0 & 0 \cr

\end{array}
\end{eqnarray}
\end{footnotesize}

\newpage
Column twentyseven to thirtyone:

\begin{footnotesize}
\begin{eqnarray}
\begin{array}{|c c c c c|}

0 & 0 & 0 & 0 & 0 \cr

0 & 0 & 0 & 0 & 0 \cr

0 & 0 & 0 & 0 & 0 \cr

0 & 0 & 0 & 0 & 0 \cr

0 & 0 & 0 & 0 & 0 \cr

\frac{1}{2}\sqrt{32}(c+\frac{k}{2}+1) & -\frac{1}{2}\sqrt{32}(c-\frac{k}{2}+1) & 0 & 0 &
\sqrt{32}(b-\frac{k}{2}) \cr

0 & 0 &-\frac{1}{2}\sqrt{32}(c+\frac{k}{2}+1)  & \frac{1}{2}\sqrt{32}(c-\frac{k}{2}+1) &
-\sqrt{32}(b+\frac{k}{2}) \cr

-\frac{1}{2}\sqrt{32}(b+\frac{k}{2}+1)& 0 & \frac{1}{2}\sqrt{32}(b-\frac{k}{2}+1) & 0 & 0
\cr

0 & \frac{1}{2}\sqrt{32}(b+\frac{k}{2}+1) & 0 & -\frac{1}{2}\sqrt{32}(b-\frac{k}{2}+1) &
0 \cr

0 & 0 & 0 & 0 & -\frac{1}{2}{\mathrm i}\sqrt{32}(a-\frac{k}{2}+1)\cr

\frac{1}{2}\sqrt{32}(a-\frac{k}{2}+1) & 0 & 0 & 0 & 0 \cr

0 & -\frac{1}{2}\sqrt{32}(a-\frac{k}{2}+1) & 0 & 0 & 0 \cr

0 & 0 & -\frac{1}{2}\sqrt{32}(a-\frac{k}{2}+1) & 0 & 0 \cr

0 & 0 & 0 & \frac{1}{2}\sqrt{32}(a-\frac{k}{2}+1) & 0 \cr

0 & 0 & 0 & 0 & 0 \cr

0 & 0 & 0 & 0 & 0 \cr

0 & 0 & 0 & 0 & 0 \cr

0 & 0 & 0 & 0 & 0 \cr

0 & 0 & 0 & 0 & 0 \cr

0 & 0 & 0 & 0 & 0 \cr

0 & 0 & 0 & 0 & 0 \cr

0 & 0 & 0 & 0 & 0 \cr

0 & 0 & 0 & 0 & 0 \cr

0 & 0 & 0 & 0 & 0 \cr

0 & 0 & 0 & 0 & 0 \cr

0 & 0 & 0 & 0 & 4k \cr

4k & 0 & 0 & 0 & 0 \cr

0 & -4k & 0 & 0 & 0 \cr

0 & 0 & -4k & 0 & 0 \cr

0 & 0 & 0 & 4k & 0 \cr

0 & 0 & 0 & 0 & 0 \cr

0 & 0 & 0 & 0 & 0 \cr

0 & 0 & 0 & 0 & 0 \cr

0 & 0 & 0 & 0 & 0 \cr

0 & 0 & 0 & 0 & 0 \cr

\end{array}
\end{eqnarray}
\end{footnotesize}

\newpage
Column thirtytwo to thirtyfive:

\begin{footnotesize}
\begin{eqnarray}
\begin{array}{|c c c c|}

 \frac{1/2}{\mathrm i}\sqrt{32}(c+\frac{k}{2}+1) & -\frac{1/2}{\mathrm i}\sqrt{32}(c-\frac{k}{2}+1)
 & \frac{1/2}{\mathrm i}\sqrt{32}(b+\frac{k}{2}+1) &  -\frac{1/2}{\mathrm i}\sqrt{32}(b-\frac{k}{2}+1) \cr

0 & 0 & \sqrt{32}(a-\frac{k}{2}) & 0\cr

0 & 0 & 0 & -\sqrt{32}(a-\frac{k}{2})  \cr

 \sqrt{32}(a-\frac{k}{2}) & 0 & 0 & 0  \cr

0 &  -\sqrt{32}(a-\frac{k}{2}) & 0 & 0  \cr

0 & 0 &  -\sqrt{32}(a+\frac{k}{2}) & 0  \cr

0 & 0 & 0 &  \sqrt{32}(a+\frac{k}{2})\cr

 -\sqrt{32}(a+\frac{k}{2})  & 0 & 0 & 0 \cr

0 & \sqrt{32}(a+\frac{k}{2}) & 0 & 0 \cr

 0 & 0 & 0 & 0 \cr

 0 & 0 & 0 & 0 \cr

 0 & 0 & 0 & 0 \cr

0 &  0 & 0 & 0 \cr

0 & 0 & 0 & 0  \cr

0 & 0 & 0 & 0 \cr

0 & 0 & -4k & 0 \cr

0 & 0 & 0 & 4k \cr

-4k & 0 & 0 & 0 \cr

 0 & 4k & 0 & 0 \cr

 0 & 0 & 0 & 0 \cr

0 & 0 & 0 & 0  \cr

0 & 0 & 0 & 0  \cr

0 & 0 & 0 &  0 \cr

0 & 0 & 0  & 0 \cr

0 & 0 &  0 & 0 \cr

0 & 0 &  0 & 0 \cr

0 & 0 &  0 & 0 \cr

0 & 0 &  0 & 0 \cr

0 & 0 &  0 & 0 \cr

0 & 0  & 0 & 0 \cr

0 &  0 & 0 & 0 \cr

0 & 0 &  0 & 0 \cr

0 & 0  & 0 & 0 \cr

0 &  0 & 0 & 0 \cr

0  & 0 & 0 & 0 \cr

\end{array}
\end{eqnarray}
\end{footnotesize}

This matrix has fifteen null eigenvalues corresponding to the longitudinal three-forms
(${\cal Y}^{(3)}=D\wedge {\cal Y}^{(2)}$), which are annihilated by $\xbox^{[111]}$
($=\,^{*}D\wedge$). These are gauge degrees of freedom that can be removed.

The remaining non-vanishing eigenvalues, associated with physical states, are:
\begin{eqnarray}
\lambda_1 =\lambda_2 &=& \ft14 \sqrt{H_0+16k+16}
\nonumber \\
\lambda_3 = \lambda_4 &=& -\ft14 \sqrt{H_0+16k+16}
\nonumber \\
\lambda_5 =\lambda_6 &=& \ft14 \sqrt{H_0-16k+16}
\nonumber \\
\lambda_7 =\lambda_8 &=& -\ft14 \sqrt{H_0-16k+16}
\nonumber\\
\lambda_9 &=& -\ft12 +\ft14\sqrt{H_0+36}
\nonumber \\
\lambda_{10} &=& -\ft12 -\ft14\sqrt{H_0+36}
\nonumber \\
\lambda_{11} = \lambda_{12} = \lambda_{13} &=& \ft12 + \ft14 \sqrt{H_0+4}
\nonumber \\
\lambda_{14} = \lambda_{15} = \lambda_{16}  &=&  \ft12 - \ft14 \sqrt{H_0+4}
\nonumber \\
\lambda_{17} &=& \ft12 +\ft14\sqrt{H_0-32k+4}
\nonumber\\
\lambda_{18} &=& \ft12 -\ft14\sqrt{H_0-32k+4}
\nonumber\\
\lambda_{19} &=& \ft12 +\ft14\sqrt{H_0+32k+4}
\nonumber\\
\lambda_{20} &=& \ft12 -\ft14\sqrt{H_0+32k+4}
\nonumber\\
\label{eigenAthreeform}
\end{eqnarray}

\subsection{The spinor}

Under the action of $H$
the 8 components of the $SO(7)$ spinor transform into the
completely reducible representation:
\begin{eqnarray}
[\frac{1}{2},\frac{1}{2},\frac{1}{2}] & \to & [-\frac{1}{2}i,0,-2i]\oplus [-\frac{1}{2}i,-i,i] \oplus [-\frac{3}{2}i,0,0] \oplus [\frac{1}{2}i,-i,-i]
\oplus [\frac{1}{2}i, i,-i] \oplus\nonumber\\ && [\frac{1}{2}i,0,2i] \oplus
  \oplus [-\frac{1}{2}i,i,i] \oplus [\frac{3}{2}i,0,0]
\end{eqnarray}
Every harmonic is then conveniently identified by the following
labels:
\begin{equation}
\begin{tabular}{|c|c|c|}
  \hline
  $\bf{J_{3}}$& $\bf{J_{3}'}$ & $\bf{J_{3}''}$ \\
  \hline\hline
  $(k-1)/2$ & $(k-1)/2$ & $(k+1)/2$  \\ \hline
  $(k-1)/2$  & $(k+1)/2$  & $(k-1)/2$  \\ \hline
  $(k-1)/2$  & $(k-1)/2$  & $(k-1)/2$  \\ \hline
  $(k-1)/2$  & $(k+1)/2$  & $(k+1)/2$  \\ \hline
  $(k+1)/2$  & $(k-1)/2$  & $(k+1)/2$  \\ \hline
  $(k+1)/2$  & $(k+1)/2$  & $(k-1)/2$  \\ \hline
  $(k+1)/2$  & $(k-1)/2$  & $(k-1)/2$  \\ \hline
  $(k+1)/2$  & $(k+1)/2$  & $(k+1)/2$  \\ \hline

  \end{tabular}
  \end{equation}
This operator acts on the $AdS_4$ fields as a eight $ \times$ eight matrix:

\newpage

Column one to three:

\begin{footnotesize}
\begin{eqnarray}
\begin{array}{| c c c|}

4\sqrt{2}(-\frac{7}{24}\sqrt{2}-\frac{1}{3}\sqrt{2}(-\frac{1}{2}+3\frac{k}{2})) & 0 &
4\sqrt{2}(-\frac{1}{2}-c+\frac{k}{2}) \cr

0 & 4\sqrt{2}(-\frac{7}{24}\sqrt{2}-\frac{1}{3}\sqrt{2}(-\frac{1}{2}+3\frac{k}{2})) &
4\sqrt{2}(\frac{1}{2}+b-\frac{k}{2}) \cr

 4\sqrt{2}(-\frac{1}{2}-c-\frac{k}{2}) &  4\sqrt{2}(\frac{1}{2}+b+\frac{k}{2}) &
4\sqrt{2}(\frac{7}{8}\sqrt{2}+\frac{1}{3}\sqrt{2}(-\frac{3}{2}+3\frac{k}{2})) \cr

 4\sqrt{2}(\frac{1}{2}+b-\frac{k}{2}) &  4\sqrt{2}(\frac{1}{2}+c-\frac{k}{2}) & 0 \cr

4\sqrt{2}(-\frac{1}{2}-a+\frac{k}{2}) & 0 & 0 \cr

0 & 4\sqrt{2}(-\frac{1}{2}-a+\frac{k}{2}) & 0 \cr

0 & 0 & 4\sqrt{2}(-\frac{1}{2}-a+\frac{k}{2}) \cr

0 & 0 & 0 \cr

\end{array}
\end{eqnarray}
\end{footnotesize}

Column four to six:
\begin{footnotesize}
\begin{eqnarray}
\begin{array}{| c c c|}

4\sqrt{2}(\frac{1}{2}+b+\frac{k}{2}) &
 4\sqrt{2}(-\frac{1}{2}-a-\frac{k}{2}) & 0 \cr

 4\sqrt{2}(\frac{1}{2}+c+\frac{k}{2}) & 0 &  4\sqrt{2}(-\frac{1}{2}-a-\frac{k}{2}) \cr

0 & 0 & 0
\cr

 4\sqrt{2}(-\frac{7}{24}\sqrt{2}+\frac{1}{3}\sqrt{2}(\frac{1}{2}+3\frac{k}{2})) &
0 & 0 \cr

0 &4\sqrt{2}(-\frac{7}{24}\sqrt{2}+\frac{1}{3}\sqrt{2}(+\frac{1}{2}+3\frac{k}{2})) & 0 \cr

0 & 0 & 4\sqrt{2}(-\frac{7}{24}\sqrt{2}+\frac{1}{3}\sqrt{2}(+\frac{1}{2}+3\frac{k}{2}))
\cr

0 & 4\sqrt{2}(\frac{1}{2}+c+\frac{k}{2}) & 4\sqrt{2}(-\frac{1}{2}-b-\frac{k}{2}) \cr

 4\sqrt{2}(-\frac{1}{2}-a+\frac{k}{2}) &
4\sqrt{2}(-\frac{1}{2}-b+\frac{k}{2}) & 4\sqrt{2}(-\frac{1}{2}-c+\frac{k}{2}) \cr

\end{array}
\end{eqnarray}
\end{footnotesize}

Column seven and eight:

\begin{footnotesize}
\begin{eqnarray}
\begin{array}{|c  c|}

 0 & 0 \cr

 0 & 0 \cr

  4\sqrt{2}(-\frac{1}{2}-a-\frac{k}{2}) & 0 \cr

 0 &  4\sqrt{2}(-\frac{1}{2}-a-\frac{k}{2}) \cr

 4\sqrt{2}(\frac{1}{2}+c-\frac{k}{2}) & 4\sqrt{2}(-\frac{1}{2}-b-\frac{k}{2}) \cr

4\sqrt{2}(-\frac{1}{2}-b+\frac{k}{2}) & 4\sqrt{2}(-\frac{1}{2}-c-\frac{k}{2})\cr

 4\sqrt{2}(-\frac{7}{24}\sqrt{2}-\frac{1}{3}\sqrt{2}(-\frac{1}{2}+3\frac{k}{2})) & 0 \cr

 0 & 4\sqrt{2}(\frac{7}{8}\sqrt{2}-\frac{1}{3}\sqrt{2}(\frac{3}{2}+3\frac{k}{2})) \cr

\end{array}
\end{eqnarray}
\end{footnotesize}

This matrix has the following eigenvalues:

\begin{eqnarray}
\lambda_1 = \lambda_2 &=& -8+\sqrt{H_0+24}
\nonumber \\
\lambda_3 = \lambda_4 &=& -8-\sqrt{H_0+24}
\nonumber \\
\lambda_5 &=& -6+\sqrt{H_{0}-16k+28}
\nonumber \\
\lambda_6 &=& -6+\sqrt{H_{0}+16k +28}
\nonumber \\
\lambda_7 &=& -6-\sqrt{H_{0}-16k +28}
\nonumber \\
\lambda_8 &=& -6-\sqrt{H_{0}+16k +28}
\nonumber \\
\label{eigenAspinor}
\end{eqnarray}

\section{Organizing the spectrum of $Osp\left( 2|4\right)\times
SU\left(2\right)\times  SU\left(2\right)'\times SU\left(2\right)''$ multiplets}

In this section we present the  $Osp\left( 2|4\right)$ multiplets that appear
compactifying $D=11$ supergravity on $AdS_{4}\times Q^{111}$. They are obtained by using
the Kaluza-Klein mass spectrum derived in the previous sections together with the mass
relations in \cite{uni,24} and the multiplet structure of the ${\cal N}=2$ multiplets. The
procedure that we use is exactly the same as that employed in \cite{m111}, to which we
refer the interested reader. We report the structure of the ${\cal N}=2$ supermultiplets
in appendix A. The irreducible supermultiplets of $Osp\left(2|4\right)$ occur into
irreducible representations of the bosonic group
\[
G'=SU\left(2\right)\times SU\left(2\right)\times SU\left(2\right)\,,\] whose
interpretation is the {\sl flavor group} on the conformal field theory side of the
correspondence and it is the {\sl gauge group} on the supergravity side. In any case the
crucial information one needs to extract from harmonic analysis is precisely the
$G^\prime$ representation assignment of the supermultiplets and the actual value of $E_0$
and $y_0$.
\par
In the following pages,  for each type  of ${\cal N}=2$ multiplet we list the  $G'$
representations through which it occurs in the spectrum. Furthermore for every multiplet,
we give   the energy and hypercharge values $E_0$ and $y_0$  of the Clifford vacuum. From
tables \ref{longgraviton}, \ref{longgravitino}, \ref{longvector}, \ref{shortgraviton},
\ref{shortgravitino}, \ref{shortvector}, \ref{hyper}, \ref{masslessgraviton},
\ref{masslessvector} it is straightforward to get the energies and the hypercharges of
all other  fields in each  multiplet.

\subsection{Long multiplets}

There are long multiplets for the following $G^{\prime}$
representations:
\begin{equation}
\left\{ J> \frac{k}{2},\ J'> \frac{k}{2},\ J''> \frac{k}{2} \right\}
\end{equation}

We have:
\begin{itemize}
\item
 one long graviton multiplet
with
\begin{equation}
h:~E_0={1\over 2}+{1\over 4}\sqrt{H_0+36},~y_0=k
\end{equation}
\item
four long gravitino multiplets (two $\chi^+$ and two $\chi^-$)
with
\begin{eqnarray}
\chi^+:~&E_0=-{1\over 2}+{1\over 4}\sqrt{H_0+16k+16},
&y_0=k-1\\
\chi^+:~&E_0=-{1\over 2}+{1\over 4}\sqrt{H_0-16k+16},
&y_0=k+1\\
\chi^-:~&E_0={3\over 2}+{1\over 4}\sqrt{H_0+16k+16},
&y_0=k-1\\
\chi^-:~&E_0={3\over 2}+{1\over 4}\sqrt{H_0-16k+16}, &y_0=k+1
\end{eqnarray}
\item
one $W$ long vector multiplet with:
\begin{equation}
W:~E_0={5\over 2}+{1\over 4}\sqrt{H_0+36},~y_0=k
\end{equation}
\item
one $A$ long vector multiplet with:
\begin{equation}
A:~E_0=-{3\over 2}+{1\over 4}\sqrt{H_0+36},~y_0=k
\end{equation}
\item
one $Z$ long vector multiplet with
\begin{equation}
Z:~E_0={1\over 2}+{1\over 4}\sqrt{H_0+4},~y_0=k
\end{equation}
\item One $Z$ long vector multiplet with
\begin{equation}
Z:~E_0={1\over 2}+{1\over 4}\sqrt{H_0+32k+4},~y_0=k
\end{equation}
\end{itemize}

\subsection{Short multiplets}

The expansion of a generic field contains only the harmonics whose
$H$- and $G$-quantum numbers are such that the $G$ representation,
decomposed under $H$, contains the $H$ representation of the
field. This fact poses some constraints on the $G$-quantum
numbers.
\par
Depending on which constraints are satisfied by a certain $G$
representation, only part of the harmonics is  present, and only
their corresponding four--dimensional fields appear in the
spectrum. Then, in the $G$ representations in which such field
disappear, there is  {\sl multiplet shortening}. In the modern
perspective of Kaluza Klein theory, the exact spectrum of the
short multiplets is crucial. Hence the importance of analyzing
this disappearance of harmonics with care. We give here all the
{\sl multiplet shortening} found  for the $Q^{111}$ manifold. As
already stressed in the introduction many of them confirm the
gauge theory predictions made in \cite{SCQ}. We find also new type
of shortening of the gravitino multiplet.
\par
We have:
\begin{itemize}
\item
One short graviton multiplet for the following $G^{\prime}$
representations:
\begin{equation}
\left\{ J=\frac{k}{2},\ J'=\frac{k}{2},\ J''=\frac{k}{2} \right\}
\end{equation}

with
\begin{equation}
E_0=k+2,~y_0=k
\end{equation}

\item
One short gravitino multiplet ($\chi^+$) for the following
$G^{\prime}$ representations:
\begin{equation}
\left\{ J=\frac{k}{2},\ J'=\frac{k}{2}+1,\ J''=\frac{k}{2}+1 \right\} \label{inosh}
\end{equation}

with
\begin{equation}
E_0=k+\frac{5}{2},~y_0=k+1
\end{equation}
All the other are obtained from (\ref{inosh}) by permuting the
role of the three $SU(2)$ groups.
\item
One short gravitino multiplet ($\chi^+$) for the following
$G^{\prime}$ representations:
\begin{equation}
\left\{ J=\frac{k}{2}-1,\ J'=\frac{k}{2},\ J''=\frac{k}{2} \right\} \label{inosho}
\end{equation}

with
\begin{equation}
E_0=k+\frac{1}{2},~y_0=k-1
\end{equation}
All the other are obtained from (\ref{inosho}) by permuting the
role of the three $SU(2)$ groups.

\item
One short gravitino multiplet ($\chi^-$) for the following $G^{\prime}$ representations:
\begin{equation}
\left\{ J=\frac{k}{2}-1,\ J'=\frac{k}{2},\ J''=\frac{k}{2} \right\} \label{inoshor}
\end{equation}

with
\begin{equation}
E_0=k+\frac{1}{2},~y_0=k-1
\end{equation}
All the other are obtained from (\ref{inoshor}) by permuting the role of the three $SU(2)$
groups.

\item
One short vector multiplet for the following $G^{\prime}$
representations:
\begin{equation}
\left\{ J=\frac{k}{2}+1,\ J'=\frac{k}{2},\ J''=\frac{k}{2} \right\} \label{vecsh}
\end{equation}

with
\begin{equation}
E_0=k+1,~y_0=k
\end{equation}
All the other are obtained from (\ref{vecsh}) by permuting the
role of the three $SU(2)$ groups.
\item
One short vector multiplet for the following $G^{\prime}$
representations:
\begin{equation}
\left\{ J=\frac{k}{2},\ J'=\frac{k}{2},\ J''=\frac{k}{2} \right\}
\end{equation}

with
\begin{equation}
E_0=k,~y_0=k
\end{equation}
\item
One hypermultiplet for the following $G^{\prime}$ representations:
\begin{equation}
\left\{ J=\frac{k}{2},\ J'=\frac{k}{2},\ J''=\frac{k}{2} \right\}
\end{equation}
with
\begin{equation}
E_0=k,~y_0=k\,\ \,\ \,\ k\geq 1
\end{equation}
\end{itemize}
\subsection{Massless multiplets}

We have:
\begin{itemize}
\item
the massless graviton multiplet in the singlet representation:

\begin{equation}
\left\{ J=0,\ J'=0,\ J''=0 \right\}
\end{equation}

with
\begin{equation}
E_0=2,~y_0=0
\end{equation}

\item
the massless vector multiplet in the adjoint representation:

\begin{equation}
\left\{ J=1,\ J'=0,\ J''=0 \right\} \label{adjvec}
\end{equation}

with
\begin{equation}
E_0=1,~y_0=0
\end{equation}
All the other are obtained from (\ref{adjvec}) by permuting the
role of the three $SU(2)$ groups.
\item
two additional massless vector multiplets in the singlet
representation:

\begin{equation}
\left\{ J=0,\ J'=0,\ J''=0 \right\}
\end{equation}

with
\begin{equation}
E_0=1,~y_0=0
\end{equation}
\end{itemize}
These two vector multiplets arises from the three form $A_{\mu i
j}$ and are due to the existence of two closed cohomology two
forms on the $Q^{111}$ manifold. These multiplets are named the
Betti multiplets.

\section{Conclusions}

In this paper I have verified that the SCFT conjectured in \cite{SCQ} to be the dual
partner of M-theory compactified on $Q^{111}$ is correct, since it reproduces the
Kaluza-Klein spectrum of such a manifold. This result was expected on general grounds but
a direct verification was so far missing. It was in my opinion important to make this
final check since it supports the geometrical picture of the dual SCFT of ${\cal N}=2$
 $AdS_4\times M^7$ compactifications based on Sasakian geometry.

\section*{Acknowledgement}

I would like to thank D. Fabbri, L. Gualtieri, A. Zaffaroni and in particular P. Fr\`e for
very useful discussions.

\newpage

\appendix

\section{Structure of the $Osp\left( 2|4\right)$ multiplets}

The structure of a supermultiplet is conveniently described by
listing the energy $E$, the spin $s$ and the hypercharge $y$ of
all the fields of the multiplet.
\par
In tables \ref {longgraviton},\ref{longgravitino}, \ref{longvector},\ref{shortgraviton},
\ref{shortgravitino},\ref{shortvector},\ref{hyper},\ref{masslessgraviton},
\ref{masslessvector} we provide such information in the first three columns. The
remaining columns provide additional information concerning the way the abstract
$Osp\left(2|4\right)$ supermultiplets are actually realized in Kaluza Klein supergravity.
Indeed for each particle state appearing in the supermultiplet we write the name of the
corresponding field in the Kaluza Klein expansion of $D=11$ supergravity to which such a
particle state contributes. For the standard expansion of linearized $D=11$ supergravity
and the conventions for the names of the $D=4$ fields we refer the reader to \cite{m111}.

The {\sl long gravitino multiplet}, satisfying $E_0>|y_0|+\ft32$,
has the structure displayed in table \ref{longgravitino}.

\begin{table}[b]
\centering
\begin{tabular}{||c|c|c|c|c||}
\hline Spin & Energy & Hypercharge & Mass ($^2$)&
Name \\
\hline \hline
$2$      & $E_0+1    $  & $y_0$     & $16(E_0+1)(E_0-2)$  & $h$ \\
$\ft32$  & $E_0+\ft32$  & $y_0-1$   & $-4E_0-4$  & $\chi^-$ \\
$\ft32$  & $E_0+\ft32$  & $y_0+1$   & $-4E_0-4$  & $\chi^-$ \\
$\ft32$  & $E_0+\ft12$  & $y_0-1$   & $4E_0-8 $  & $\chi^+$ \\
$\ft32$  & $E_0+\ft12$  & $y_0+1$   & $4E_0-8 $  & $\chi^+$ \\
$1$      & $E_0+2    $  & $y_0$     & $16E_0(E_0+1)$    & $W$ \\
$1$      & $E_0+1    $  & $y_0-2$   & $16E_0(E_0-1)$    & $Z$ \\
$1$      & $E_0+1    $  & $y_0+2$   & $16E_0(E_0-1)$    & $Z$ \\
$1$      & $E_0+1    $  & $y_0$     & $16E_0(E_0-1)$    & $Z$ \\
$1$      & $E_0+1    $  & $y_0$     & $16E_0(E_0-1)$    & $Z$ \\
$1$      & $E_0      $  & $y_0$     & $16(E_0-1)(E_0-2)$& $A$  \\
$\ft12$  & $E_0+\ft32$  & $y_0-1$   & $4E_0$     &  $\lambda_T$  \\
$\ft12$  & $E_0+\ft32$  & $y_0+1$   & $4E_0$     &  $\lambda_T$ \\
$\ft12$  & $E_0+\ft12$  & $y_0-1$   & $-4E_0+4$  &  $\lambda_T$ \\
$\ft12$  & $E_0+\ft12$  & $y_0+1$   & $-4E_0+4$  &  $\lambda_T$ \\
$0$      & $E_0+1    $  & $y_0$     & $16E_0(E_0-1)  $  & $\phi$ \\
\hline
\end{tabular}\\[.13in]
\caption{${\cal N}=2$ long graviton multiplet}
\label{longgraviton}
\end{table}

\begin{table}[b]
\centering
\begin{small}
\begin{tabular}{||c|c|c|c|c|c|c||}
\hline Spin & Energy & Hypercharge & Mass ($^2$)& Name & Mass
($^2$)&
name \\
\hline \hline $\ft32$  & $E_0+1$      & $y_0$   & $4E_0-6$  &
$\chi^+$
				& $-4E_0-2$ & $\chi^-$  \\
$1$      & $E_0+\ft32$  & $y_0-1$ & $16(E_0-\ft12)(E_0+\ft12)$ &
$Z$
				& $16(E_0-\ft12)(E_0+\ft12)$ & $W$ \\
$1$      & $E_0+\ft32$  & $y_0+1$ & $16(E_0-\ft12)(E_0+\ft12)$ &
$Z$
				& $16(E_0-\ft12)(E_0+\ft12)$ & $W$ \\
$1$      & $E_0+\ft12$  & $y_0-1$ & $16(E_0-\ft32)(E_0-\ft12)$ &
$A$
				& $16(E_0-\ft32)(E_0-\ft12)$ & $Z$ \\
$1$      & $E_0+\ft12$  & $y_0+1$ & $16(E_0-\ft32)(E_0-\ft12)$ &
$A$
				& $16(E_0-\ft32)(E_0-\ft12)$ & $Z$ \\
$\ft12$  & $E_0+2$      & $y_0$   & $4E_0+2$  &  $\lambda_T$
				& $-4E_0-2$ &  $\lambda_L$ \\
$\ft12$  & $E_0+1$      & $y_0-2$ & $-4E_0+2$ &  $\lambda_T$
				& $-4E_0-2$ &  $\lambda_T$ \\
$\ft12$  & $E_0+1$      & $y_0$   & $-4E_0+2$ &  $\lambda_T$
				& $4E_0-2$  &  $\lambda_T$  \\
$\ft12$  & $E_0+1$      & $y_0+2$ & $-4E_0+2$ &  $\lambda_T$
				& $4E_0-2$  &  $\lambda_T$ \\
$\ft12$  & $E_0+1$      & $y_0$   & $-4E_0+2$ &  $\lambda_T$
				& $4E_0-2$  &  $\lambda_T$  \\
$\ft12$  & $E_0$        & $y_0$   & $4E_0-6$  &  $\lambda_L$
				& $-4E_0+6$ &  $\lambda_T$ \\
$0$      & $E_0+\ft32$  & $y_0-1$ & $16(E_0-\ft12)(E_0+\ft12)$  &
$\phi$
				& $16(E_0-\ft12)(E_0+\ft12)$  & $\pi$ \\
$0$      & $E_0+\ft32$  & $y_0+1$ & $16(E_0-\ft12)(E_0+\ft12)$  &
$\phi$
				& $16(E_0-\ft12)(E_0+\ft12)$  & $\pi$ \\
$0$      & $E_0+\ft12$  & $y_0-1$ & $16(E_0-\ft32)(E_0-\ft12)$  &
$\pi$
				& $16(E_0-\ft32)(E_0-\ft12)$  & $\phi$ \\
$0$      & $E_0+\ft12$  & $y_0+1$ & $16(E_0-\ft32)(E_0-\ft12)$  &
$\pi$
				& $16(E_0-\ft32)(E_0-\ft12)$  & $\phi$ \\
\hline
\end{tabular}\\[.13in]
\end{small}
\caption{${\cal N}=2$ long gravitino multiplets $\chi^+$ and
$\chi^-$} \label{longgravitino}
\end{table}

The {\sl long vector multiplet}, satisfying $E_0>|y_0|+1$, has
the structure displayed in table \ref{longvector}.
\begin{table}[b]
\centering
\begin{footnotesize}
\begin{tabular}{||c|c|c|c|c|c|c|c||}
\hline Spin & Energy & Hypercharge & Mass ($^2$)& Name & Name &
Mass ($^2$)&
Name \\
\hline \hline $1$      & $E_0+1$      & $y_0$   & $16E_0(E_0-1)$
& $A$
				 & $W$     & $16E_0(E_0-1)$  & $Z$      \\
$\ft12$  & $E_0+\ft32$  & $y_0-1$ & $-4E_0$  &   $\lambda_T$
				 &   $\lambda_L$  & $4E_0$   &   $\lambda_T$   \\
$\ft12$  & $E_0+\ft32$  & $y_0+1$ & $-4E_0$  &   $\lambda_T$
				  &   $\lambda_L$  & $4E_0$   &   $\lambda_T$   \\
$\ft12$  & $E_0+\ft12$  & $y_0-1$ & $4E_0-4$ &   $\lambda_L$
				 &   $\lambda_T$   & $-4E_0+4$&   $\lambda_T$ \\
$\ft12$  & $E_0+\ft12$  & $y_0+1$ & $4E_0-4$ &   $\lambda_L$
				 &   $\lambda_T$   & $-4E_0+4$&   $\lambda_T$  \\
$0$      & $E_0+2$      & $y_0$   & $16E_0(E_0+1)$  & $\phi$
				 & $\Sigma$  & $16E_0(E_0+1)$  & $\pi$  \\
$0$      & $E_0+1$      & $y_0-2$ & $16E_0(E_0-1)$  & $\pi$
				  & $\pi$   & $16E_0(E_0-1)$  & $\phi$   \\
$0$      & $E_0+1$      & $y_0+2$ & $16E_0(E_0-1)$  & $\pi$
				  & $\pi$   & $16E_0(E_0-1)$  & $\phi$   \\
$0$      & $E_0+1$      & $y_0$   & $16E_0(E_0-1)$  & $\pi$
				  & $\pi$    & $16E_0(E_0-1)$  & $\phi$   \\
$0$      & $E_0$        & $y_0$   & $16(E_0-2)(E_0-1)$  & $S$
				  & $\phi$  & $16(E_0-2)(E_0-1)$  & $\pi$  \\
\hline
\end{tabular}\\[.13in]
\end{footnotesize}
\caption{${\cal N}=2$ long vector multiplets $A$,$W$ and $Z$}
\label{longvector}
\end{table}
Also the long vector multiplet has different realizations from the
Kaluza Klein viewpoint.
\par
The short multiplets are of two kinds: the {\sl short graviton,
gravitino} and {\sl vector  multiplets}, that saturate the bound
\begin{equation}
E_0=|y_0|+s_0+1 \label{shortcond}
\end{equation}
and the {\sl hypermultiplets} (spin $1/2$ multiplets), that
saturate the other bound
\begin{equation}
E_0=|y_0|~~{\rm with}~~|y_0|\ge{1\over 2}. \label{masslesscond}
\end{equation}
\par
The {\sl short graviton multiplet}, satisfying $E_0=|y_0|+2$, has
the structure displayed in table \ref{shortgraviton}.
\begin{table}[b]
\centering
\begin{footnotesize}
\begin{tabular}{||c|c|c|c|c||}
\hline Spin & Energy & Hypercharge & Mass ($^2$)&
Name \\
\hline \hline
$2$      & $y_0+3$        & $y_0$     & $16y_0(y_0+3)$  & $h$        \\
$\ft32$  & $y_0+\ft72$    & $y_0-1$& $-4y_0-12$  & $\chi^-$      \\
$\ft32$  & $y_0+\ft52$    & $y_0+1$   & $4y_0 $  & $\chi^+$             \\
$\ft32$  & $y_0+\ft52$    & $y_0-1$   & $4y_0 $  & $\chi^+$             \\
$1$      & $y_0+3    $    & $y_0-2$& $16(y_0+2)(y_0+1)$    & $Z$        \\
$1$      & $y_0+3    $    & $y_0$     & $16(y_0+2)(y_0+1)$    & $Z$     \\
$1$      & $y_0+2      $  & $y_0$     & $16y_0(y_0+1)$& $A$         \\
$\ft12$  & $y_0+\ft52$  & $y_0-1$& $-4y_0-4$  &  $\lambda_T$    \\
\hline
\end{tabular}\\[.13in]
\end{footnotesize}
\caption{${\cal N}=2$ short graviton multiplet with positive
hypercharge $y_0>0$} \label{shortgraviton}
\end{table}
\par
The {\sl short gravitino multiplet}, satisfying
$E_0=|y_0|+\ft32$, has the structure displayed in table
\ref{shortgravitino}.
\begin{table}[b]
\centering
\begin{footnotesize}
\begin{tabular}{||c|c|c|c|c||}
\hline Spin & Energy & Hypercharge & Mass ($^2$)&
Name \\
\hline \hline
$\ft32$  & $y_0+\ft52$    & $y_0$         & $4y_0$                  & $\chi^+$  \\
$1$      & $y_0+3$        & $y_0-1$  & $16(y_0+1)(y_0+2)$      & $Z$    \\
$1$      & $y_0+2$        & $y_0+1$       & $16y_0(y_0+1)$       & $A$  \\
$1$      & $y_0+2$        & $y_0-1$       & $16y_0(y_0+1)$       & $A$  \\
$\ft12$   & $y_0+\ft52$   & $y_0$         & $-4y_0-4$            &  $\lambda_T$ \\
$\ft12$  & $y_0+\ft52$    & $y_0-2$         & $-4y_0-4$            &  $\lambda_T$   \\
$\ft12$  & $y_0+\ft32$    & $y_0$         & $4y_0$                  &  $\lambda_L$  \\
$0$       &  $y_0+3$   & $y_0\pm 1$     & $16(y_0+1)(y_0+2)$ &  $\phi$  \\
\hline
\end{tabular}\\[.13in]
\end{footnotesize}
\caption{${\cal N}=2$ short gravitino multiplet $\chi^+$ with
positive hypercharge $y_0 >0$} \label{shortgravitino}
\end{table}
\par
The {\sl short vector multiplet}, satisfying $E_0=|y_0|+1$, has the structure displayed
in table \ref{shortvector}.
\par
We must stress that the multiplets displayed in tables
\ref{shortgraviton}, \ref{shortgravitino}, \ref{shortvector} are
only half of the story, since they can be viewed as the BPS
states where $E_0 = y_0 +s_0 +1$ and $y_0 >0 $. In addition one
has also the anti BPS states. These are the short multiplets
where $E_0 = -y_0 +s_0 +1$ with $y_0 <0$.  The structure of these
anti short multiplets can be easily read off  from tables
\ref{shortgraviton}, \ref{shortgravitino}, \ref{shortvector} by
reversing the sign of all hypercharges.
\par
The {\sl hypermultiplet}, satisfying $E_0=|y_0|\ge{1\over 2}$,
has the structure displayed in table \ref{hyper}.
\begin{table}[b]
\centering
\begin{footnotesize}
\begin{tabular}{||c|c|c|c|c||}
\hline Spin & Energy & Hypercharge & Mass ($^2$)&
Name  \\
\hline \hline
$1$      & $y_0+2$      & $y_0$      & $16y_0(y_0+1)$   & $A$       \\
$\ft12$  & $y_0+\ft52$  & $y_0\pm 1$ & $-4y_0-4$    & $\lambda_T$   \\
$\ft12$  & $y_0+\ft32$  & $y_0+1$    & $4y_0$       & $\lambda_L$   \\
$\ft12$  & $y_0+\ft32$  & $y_0-1$    & $4y_0$       & $\lambda_L$   \\
$0$      & $y_0+2$      & $y_0-2$ & $16y_0(y_0+1)$  & $\pi$     \\
$0$      & $y_0+2$      & $y_0$      & $16y_0(y_0+1)$   & $\pi$         \\
$0$      & $y_0+1$      & $y_0$      & $16y_0(y_0-1)$   & $S$       \\
\hline
\end{tabular}\\[.13in]
\end{footnotesize}
\caption{${\cal N}=2$ short vector multiplet $A$ with positive
hypercharge $y_0 >0$} \label{shortvector}
\end{table}
\begin{table}[b]
\centering
\begin{footnotesize}
\begin{tabular}{||c|c|c|c|c||}
\hline Spin & Energy & Hypercharge & Mass ($^2$)&
Name  \\
\hline \hline
$\ft12$  & $y_0+\ft12$  & $y_0-1$    & $4y_0-4$       & $\lambda_L$   \\
$0$      & $y_0+1$      & $y_0-2$    & $16y_0(y_0-1)$   & $\pi$     \\
$0$      & $y_0$        & $y_0$      & $16(y_0-2)(y_0-1)$   & $S$       \\
\hline \hline
$\ft12$  & $y_0+\ft12$  & $-y_0+1$    & $4y_0-4$      & $\lambda_L$   \\
$0$      & $y_0+1$      & $-y_0+2$    & $16y_0(y_0-1)$   & $\pi$     \\
$0$      & $y_0$        & $-y_0$      & $16(y_0-1)(y_0-2)$   & $S$       \\
\hline
\end{tabular}\\[.13in]
\end{footnotesize}
\caption{${\cal N}=2$ hypermultiplet, $y_0>0$} \label{hyper}
\end{table}
\par
The massless multiplets are either  short graviton or short vector
multiplets satisfying the further condition
\begin{equation}
E_0=s_0+1~~~{\rm equivalent~to}~~~y_0=0.
\end{equation}
\par
The {\sl massless graviton multiplet}, satisfying $E_0=2~~y_0=0$,
has the structure displayed in table \ref{masslessgraviton}.
\begin{table}[b]
\centering
\begin{footnotesize}
\begin{tabular}{||c|c|c|c|c||}
\hline Spin & Energy & Hypercharge & Mass ($^2$)&
Name \\
\hline \hline
$2$      & $3$        & $0$     & $0$  & $h$     \\
$\ft32$  & $\ft52$    & $-1$   & $0$  & $\chi^+$\\
$\ft32$  & $\ft52$    & $+1$   & $0$  & $\chi^+$\\
$1$      & $2$        & $0$     & $0$  & $A$     \\
\hline
\end{tabular}\\[.13in]
\end{footnotesize}
\caption{${\cal N}=2$ massless graviton multiplet}
\label{masslessgraviton}
\end{table}
\par
The {\sl massless vector multiplet}, satisfying $E_0=1~~y_0=0$,
has the structure displayed in table \ref{masslessvector}.
\begin{table}[b]
\centering
\begin{footnotesize}
\begin{tabular}{||c|c|c|c|c|c|c||}
\hline Spin & Energy & Hypercharge & Mass ($^2$)& Name & Mass
($^2$)&
Name \\
\hline \hline $1$      & $2$      & $0$      & $0$  & $A$
					   & $0$  & $Z$         \\
$\ft12$  & $\ft32$  & $-1$    & $0$  & $\lambda_L$
				   & $0$  & $\lambda_T$ \\
$\ft12$  & $\ft32$  & $+1$    & $0$  & $\lambda_L$
				   & $0$  & $\lambda_T$ \\
$0$      & $2$      & $0$ & $0$  & $\pi$
				   & $0$ & $\phi$      \\
$0$      & $1$        & $0$      & $0$  & $S$
				   & $0$  & $\pi$   \\
\hline
\end{tabular}\\[.13in]
\end{footnotesize}
\caption{${\cal N}=2$ massless vector multiplets $A$ and $Z$}
\label{masslessvector}
\end{table}
These are the ${\cal N}=2$ supermultiplets in anti de Sitter
space that can occur  in Kaluza Klein supergravity.

\end{document}